\documentclass[sigconf,screen]{acmart}

\usepackage{graphicx}
\usepackage{setspace}               
\usepackage{multicol}               
\usepackage{xcolor}
\usepackage{caption}
\usepackage{subcaption}
\usepackage{csquotes}

\usepackage[utf8]{inputenc}
\usepackage{url}
\usepackage{hyperref}
\usepackage{amsfonts}
\usepackage{float} 
\usepackage{listings}
\usepackage{color}
\usepackage{fancyhdr} 
\usepackage{bibentry} 
\usepackage{enumitem}
\usepackage{multirow}
\usepackage{geometry}
\usepackage{dblfloatfix}

\usepackage{array}

\usepackage{dirtytalk}

\usepackage{soul}

\definecolor{edit}{HTML}{000000}
\definecolor{minor}{HTML}{000000}

\AtBeginDocument{%
  \providecommand\BibTeX{{%
    \normalfont B\kern-0.5em{\scshape i\kern-0.25em b}\kern-0.8em\TeX}}}

\copyrightyear{2025} 
\acmYear{2025} 
\setcopyright{cc}
\setcctype{by}
\acmConference[CUI '25]{Proceedings of the 7th ACM Conference on Conversational User Interfaces}{July 8--10, 2025}{Waterloo, ON, Canada}
\acmBooktitle{Proceedings of the 7th ACM Conference on Conversational User Interfaces (CUI '25), July 8--10, 2025, Waterloo, ON, Canada}
\acmDOI{10.1145/3719160.3736617}
\acmISBN{979-8-4007-1527-3/2025/07}

\begin{document}

\title[The Impact of a Chatbot's Ephemerality-Framing on Self-Disclosure Perceptions]{The Impact of a Chatbot's Ephemerality-Framing on Self-Disclosure Perceptions}

\author{Samuel Rhys Cox}
\email{srcox@cs.aau.dk}
\orcid{0000-0002-4558-6610}
\affiliation{%
  \institution{Aalborg University}
  \city{Aalborg}
  \country{Denmark}
}

\author{Rune M{\o}berg Jacobsen}
\email{runemj@cs.aau.dk}
\orcid{0000-0002-1877-1845}
\affiliation{%
  \institution{Aalborg University}
  \city{Aalborg}
  \country{Denmark}
}

\author{Niels van Berkel}
\email{nielsvanberkel@cs.aau.dk}
\orcid{0000-0001-5106-7692}
\affiliation{%
  \institution{Aalborg University}
  \city{Aalborg}
  \country{Denmark}
}

\begin{abstract}





Self-disclosure, the sharing of one's thoughts and feelings, is affected by the perceived relationship between individuals. While chatbots are increasingly used for self-disclosure, the impact of a chatbot's framing on users' self-disclosure remains under-explored. We investigated how a chatbot’s description of its relationship with users, particularly in terms of ephemerality, affects self-disclosure. Specifically, we compared a \textsc{Familiar} chatbot, presenting itself as a companion remembering past interactions, with a \textsc{Stranger} chatbot, presenting itself as a new, unacquainted entity in each conversation. 
In a mixed factorial design, participants engaged with either the \textsc{Familiar} or \textsc{Stranger} chatbot in two sessions across two days, with one conversation focusing on \textsc{Emotional}- and another \textsc{Factual}-disclosure.
When \textsc{Emotional}-disclosure was sought in the first chatting session, \textsc{Stranger}-condition participants felt more comfortable self-disclosing. However, when \textsc{Factual}-disclosure was sought first, these differences were replaced by more enjoyment among \textsc{Familiar}-condition participants. 
Qualitative findings showed \textsc{Stranger} afforded anonymity and reduced judgement, whereas \textsc{Familiar} sometimes felt intrusive unless rapport was built via low-risk \textsc{Factual}-disclosure.

\end{abstract}

\ccsdesc[500]{Human-centered computing~Empirical studies in HCI}

\keywords{Chatbots, Self-Disclosure, Emotional Disclosure, Ephemerality, Temporality}

\begin{teaserfigure}
\centering
 \includegraphics[width=1\textwidth]{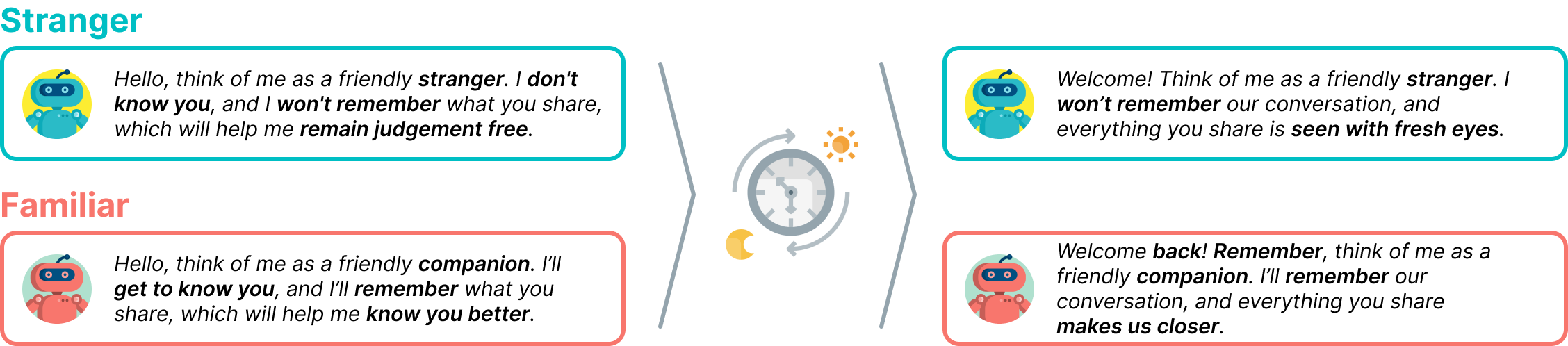}
 \caption{Across two chatting sessions, users talked to either the \textsc{Familiar} chatbot (framing itself as a lasting companion), or the \textsc{Stranger} chatbot (framing itself as a new, unacquainted interlocutor each time). We then measured the effect of framing on self-disclosure perceptions.}
 \Description{}
 \label{fig:Experiment-Flow}
\end{teaserfigure}


\maketitle


\section{Introduction}


Human relationships are complex and diverse, with different relationships affecting how we interact, talk, and share with others.
For example, we may feel comfortable self-disclosing our thoughts, feelings, and emotions to friends or family due to established intimacy and trust~\cite{altman1973socialpenetrationtheory}.
Conversely, we may also self-disclose to complete strangers if we perceive the interaction as fleeting and ephemeral, with no long-term consequences~\cite{altman1973socialpenetrationtheory, derlega1977privacy,perlman1987revelation}, for example when confiding in a stranger on the train. 
These decisions involve a risk-reward calculus, weighing the potential benefits and risks of sharing information. 
Intimate relations offer a sense of security and trust built over time, while the anonymity and lack of long-term consequences in interactions with strangers can provide catharsis and a third-party perspective. 
This dichotomy highlights the role of \textbf{\textit{ephemerality}}, the temporal nature of an interaction, in shaping our willingness to self-disclose.

Investigations of the ephemerality of relationships, interactions, and self-disclosure have recently taken place in the digital space.
For example, in social media communication~\cite{bernstein20114chan,duradoni2021stranger,misoch2015stranger,yao2024s,bayer2016sharing} a ``\textit{stranger on the Internet}'' phenomena has emerged~\cite{duradoni2021stranger,misoch2015stranger} in which people may feel more comfortable disclosing to strangers online.
Alongside this, conversational agents (CAs) are increasingly used as a means for people to share their personal thoughts and feelings~\cite{jo2024understanding,park2019designing,seo2023chacha}, and large language models (LLMs) are driving the popularity of CAs into the public domain for a wide variety of tasks.
In relation to the concept of ephemerality, ChatGPT (one of the major LLM-driven CAs) now offers memory across chatting sessions~\cite{ChatGPT-memory-article} (to offer a more personalised and continuous interaction experience), as well as a ``\textit{temporary chat}'' feature where memory of a chatting session is not kept upon its conclusion (to offer user privacy and separation of conversation contexts).
These dual modes of interaction, integrating persistent or ephemeral memory, offer distinct advantages. Yet, it remains unclear how these different approaches affect user comfort and perceptions towards self-disclosure to CAs.
That is to say, while the temporal nature of interactions has been investigated in human-human interactions~\cite{altman1973socialpenetrationtheory}, the convention within human-computer interactions has been to design CAs that emphasise the importance of building long-term relationships with users~\cite{skjuve2023longitudinal}.



Therefore, we investigate the effect of how the ephemerality of an interaction is framed when users are talking to a chatbot, and how it impacts a user's perceptions related to self-disclosure.
Specifically, we compared a \textsc{Familiar} chatbot, which presented itself as a companion that would remember past interactions, with a \textsc{Stranger} chatbot, which presented itself as a new, unacquainted entity in each conversation.
Following a 2 $\times$ 2 mixed factorial design, participants interacted with either the \textsc{Familiar} or \textsc{Stranger} chatbot in two sessions over two days, with one conversation centred on \textsc{Emotional}-disclosure (i.e., loneliness and social connections) and the other on \textsc{Factual}-disclosure (i.e., hobbies and interests).

From our user study, we found evidence (both quantitative and qualitative) that the ephemerality-framing of a chatbot can impact perceptions surrounding self-disclosure. 
This was in addition to effects due to the order of disclosure type sought in each chatting session. Specifically, if more sensitive \textsc{Emotional}-disclosure was sought in the first chatting session, participants in the \textsc{Stranger} condition reported feeling more comfortable.
However, if \textsc{Factual}-disclosure was sought in the first chatting session, differences in comfort were no longer significant, and participants in the \textsc{Familiar} condition reported greater levels of enjoyment and desire to continue.
Qualitative findings showed that the \textsc{Stranger} framing afforded anonymity and reduced judgement, whereas the \textsc{Familiar} framing sometimes felt intrusive (unless rapport was first developed via low-risk \textsc{Factual}-disclosure).
In addition, users’ beliefs regarding the chatbot’s emotional capacity both facilitated and, for some, discouraged deeper self-disclosure.


\section{Related Work}


\subsection{Factors Influencing Self-Disclosure: Offline and Online Contexts}
\label{sec:RW_self-disclosure}


Self-disclosure is the act of revealing personal information to others, such as one's thoughts, feelings, and opinions~\cite{archer1980effects}.
Mental health experts often recommend self-disclosure because it can reduce stress, promote self-reflection, strengthen social bonds, and encourage social support~\cite{greene2006self}.
Despite these advantages, people may hesitate to self-disclose due to fears of negative outcomes, such as being judged, stereotyped, or perceiving that the intended recipient is not receptive to their disclosure~\cite{vogel2003seek}.
As posited by Social Penetration Theory, this is part of a risk-reward calculus where people weigh up both the expected risks and benefits from self-disclosure~\cite{altman1973socialpenetrationtheory}. 
Furthermore, the willingness to self-disclose and the depth of disclosure can vary significantly depending on the nature of a relationship. 
Relationships develop in stages, moving from superficial to more intimate levels of disclosure as trust and familiarity increase, and as people's relationships develop it is \textit{generally} expected that their self-disclosure will increase~\cite{ruppel2015use,altman1973socialpenetrationtheory}.

In this process, the \textit{type of relationship} plays a critical role in determining both the content and frequency of self-disclosure.
For example, while we may be more selective and strategic in our self-disclosure to work colleagues (with people moderating between self-disclosure to increase workplace camaraderie, and limiting depth of disclosure to maintain professionalism), we may disclose more to friends due to emotional closeness and mutual trust that typically characterise these relationships~\cite{altman1973socialpenetrationtheory}.
Meanwhile, when interacting with strangers, the likelihood of self-disclosure is typically lower with people being more cautious in disclosing personal information due to lack of trust and uncertainty regarding how disclosures will be received typically outweighing potential benefits.
However, there are exceptions in contexts where perceived anonymity from strangers provides a sense of security when disclosing~\cite{frye2010trust,ma2016anonymity,duradoni2021stranger,misoch2015stranger}.
For example, the ``\textit{stranger on the train}'' phenomenon refers to the observation that people may feel comfortable disclosing to complete strangers, particularly in transient or ephemeral situations where the likelihood of a lasting relationship is minimal~\cite{altman1973socialpenetrationtheory, derlega1977privacy,perlman1987revelation}. In such encounters, the fleeting nature reduces perceived social risks, encouraging more open self-disclosure without fear of long-term consequences or judgement.

A similar effect, the ``\textit{stranger on the Internet}'' phenomenon, has been observed in online communities whereby people may disclose more in anonymous or pseudo-anonymous contexts~\cite{duradoni2021stranger,misoch2015stranger}. 
The extent to which people self-disclose in such online contexts has been shown to be affected by both the perceptions of anonymity~\cite{ma2016anonymity}, the relationship to and intimacy towards those in the online social circle (e.g., social media connections)~\cite{andalibi2019happens,yao2024s}, and the user's perceived ephemerality of the disclosure~\cite{yao2024s,bayer2016sharing}.
Specifically, Ma et al.\ found that anonymity on social media increases the baseline of self-disclosure, especially when sharing negative-valence content such as bad experiences in romantic relationships, which are often perceived as more intimate~\cite{ma2016anonymity}.
Moreover, people are more willing to share personal information if the communication tool is seen as more privacy-preserving~\cite{frye2010trust}.
When people believe their content will not persist, as seen on platforms designed for temporary sharing, they feel less concerned with self-presentation and are more likely to engage in unfiltered and authentic interactions~\cite{bayer2016sharing, yao2024s}. 
On from this, Yao et al.\ found that people sharing over a more ephemeral source (i.e., Instagram Stories) self-disclosed more due to both the impermanence of the post, as well as perceiving the audience of said post as being more intimate and close to themselves.
Furthermore, Ma et al.\ found that when users posted ephemeral or anonymous questions on a question asking website, it led to reduced social costs in asking questions~\cite{ma2019effects}.

\begin{figure*}[htbp]
  \centering
  \includegraphics[width=0.8\textwidth]{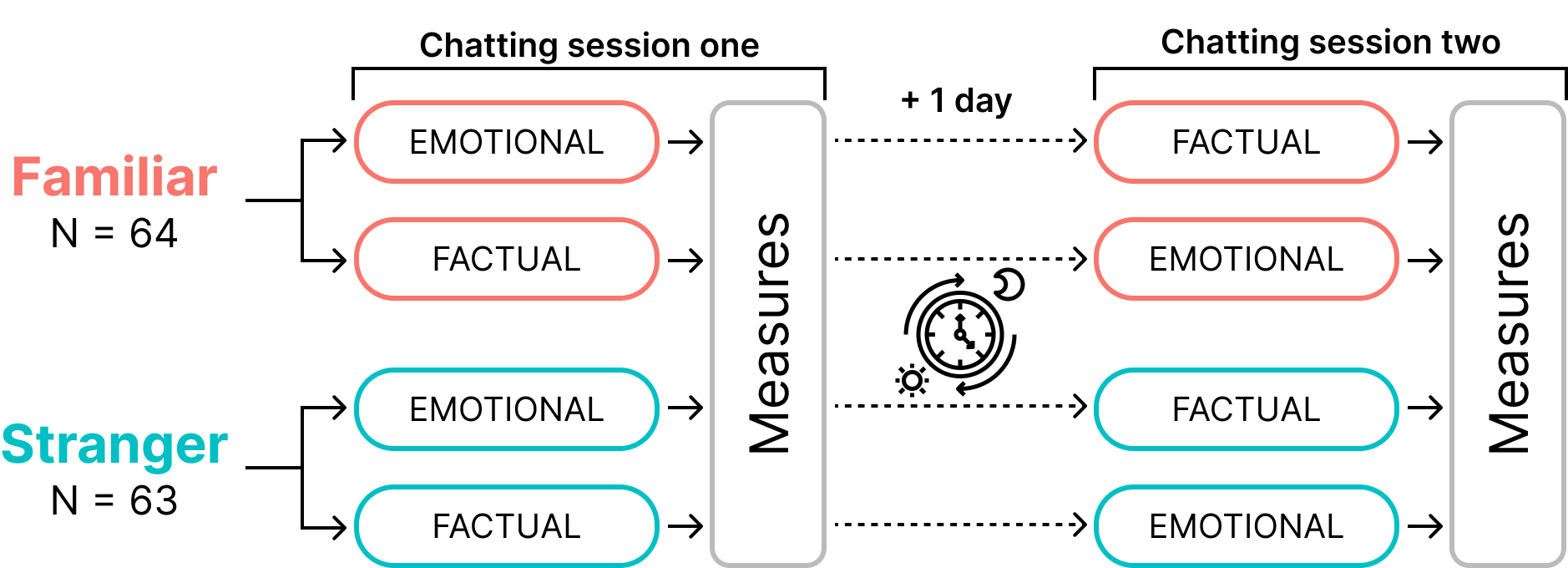}
  \caption{The experiment flow for user studies. Participants talked to either the \textsc{Familiar} or \textsc{Stranger} chatbot across two separate chatting sessions, with one session eliciting \textsc{Emotional}-disclosure and one \textsc{Factual}-disclosure.}
  \Description{}
  \label{fig:Experiment-Flow}
\end{figure*}

Beyond the relationship of those we are self-disclosing to, the \textit{type of information} that is being disclosed will affect both depth and comfort when disclosing~\cite{laurenceau1998intimacy,gomez2023sensitive}.
Self-disclosure can be divided into both factual- or emotional-disclosure~\cite{laurenceau1998intimacy}.
\textit{Factual-disclosure} is the process of revealing objective or impersonal information, such as one’s hobbies, preferences, or factual details about one’s life (e.g., favourite books, food preferences, or daily activities). 
This type of disclosure is generally perceived as lower-risk, as it involves less sensitive or intimate content~\cite{rubin1978friendship}. 
In contrast, \textit{emotional-disclosure} refers to the sharing of personal feelings, thoughts, or experiences that reflect one's emotional state, such as discussing loneliness, relationships, or personal challenges. 
Emotional-disclosure is often more intimate, requiring greater vulnerability, and as a result, individuals may feel more hesitant or uncomfortable sharing this type of information depending on the context and the perceived safety of the interaction~\cite{vogel2003seek}.

In this paper, we combine these facets to investigate the role of \textit{ephemerality} and \textit{disclosure type} when people self-disclose to a chatbot. In the next section, we therefore overview work related to self-disclosure to chatbots.



\subsection{Conversational Agents and Self-Disclosure}

Conversational agents are being increasingly used for wellbeing support where people can discuss their potential problems~\cite{park2021designing,park2019designing,seo2023chacha,jin2024effects,Wester2024MoralAgency}, such as graduate students discussing their stress~\cite{park2019designing}.
It has also been found that (similarly to human relationships) self-disclosure is important in forming human-CA relationships~\cite{skjuve2023longitudinal,skjuve2022longitudinal,croes2021can,joinson2001self}, and self-disclosure to a chatbot \textit{may} be equally beneficial to self-disclosure to a human~\cite{ho2018psychological}.
On from this, prior work has investigated how adapting social cues of CAs can increase self-disclosure in health~\cite{cox2022does} and wellbeing contexts~\cite{lee2020designing,lee2023user}.
For example, it has been found that a more formal conversational style may encourage self-disclosure within a healthcare context~\cite{cox2022does}, and that CA use of self-disclosure can encourage reciprocal self-disclosure from users themselves~\cite{lee2020designing,moon2000intimate}.
Additionally, Zhang et al.\ found that people had more willingness to disclose to an agent that adopted the role of a mentor compared to that of a servant~\cite{zhang2023tools}.


Recent studies have also investigated the impact of a CA's memory abilities and data management on user perceptions and self-disclosure~\cite{jo2024understanding,cox2023comparing,joincorporating,liu2024compeer}.
Cox et al.\ found that when a CA explicitly references a user's prior utterances (that are stored in memory from previous chatting sessions), it increases feelings of engagement but raises privacy concerns~\cite{cox2023comparing}.
Similarly, Jo et al.\ investigated the use of long-term memory in LLM-powered CAs, and found that while it could increase feelings of familiarity and foster sharing, there is also a tension raised with increased privacy concerns~\cite{jo2024understanding}. 
In relation to these privacy concerns, prior work has focused on giving users greater levels of control over a CA's memory, such as choices regarding data retention by the agent~\cite{lau2018alexa,phinnemore2023creepy,thudt2015visual}.
However, recent work has found that some users are already disclosing their sensitive personal information to LLMs, in this case to ChatGPT~\cite{zhang2024s}. This effect could be exacerbated over longitudinal use of CAs, where users may provide greater depth and breadth of disclosure as longitudinal interaction takes place~\cite{skjuve2023longitudinal}.

This prior work has therefore studied the privacy and self-dis\-closure implications of CAs that possess the ability to store and reference previous user utterances.
However, and as outlined in Section \ref{sec:RW_self-disclosure}, the perceived relationship with an interlocutor has ranging impacts on someone's self-disclosure that can be present in both human-human and human-computer interactions.
On from this, we explore \textit{how} the relationship of a CA agent is framed in terms of the ephemerality of the interaction (a factor that has strong implications for self-disclosure behaviour and perceptions~\cite{altman1973socialpenetrationtheory,duradoni2021stranger,misoch2015stranger}) as well as the interactional effect with the type of self-disclosure that is elicited by a CA interlocutor.

\begin{table*}[htbp]
\caption{Conversational features present in the \textsc{Stranger} and \textsc{Familiar} conditions.}
\label{tab:conversational-features}
\begin{tabular}{p{0.21\linewidth}p{0.37\linewidth}p{0.37\linewidth}}
\toprule
  & \textsc{Stranger} & \textsc{Familiar} \\
\midrule
 Metaphor of relationship & 
 The chatbot describes itself as a ``\textit{friendly stranger}'' who does not know the user. & 
 The chatbot describes itself as a ``\textit{friendly companion}'' who will get to know the user more as they interact. \\ 
 Memory of interactions (verbal) & 
 The chatbot states that it \textit{will not remember} the chatting session upon completion, does not have knowledge or familiarity of the user, and therefore does not have prior knowledge of interactions with the user outside of the current chatting session.
 & The chatbot states that it \textit{will remember} the chatting session, will get to know the user and grow familiar with them, and that (if applicable) it has prior knowledge of interactions. \\
 Memory of interactions (visual) & 
 Each chatting session is independent, and the content of previous chatting sessions is not shown to users.
 & Contents of the previous chatting session are displayed above subsequent chatting sessions. \\
 Purported benefits (derived from SPT~\cite{altman1973socialpenetrationtheory}) &
 Benefits pertinent to ephemeral conversations are stated (i.e., ``[...] \textit{will help me stay free from prior judgement}''). &
 Benefits pertinent to persistent conversations are stated (i.e., ``[...] \textit{will help me get to know you better}''). \\
 Emphasising (im)permanence &
 The chatbot's language emphasises the \textbf{ephemeral} nature of the conversation, and language makes no reference or allusion to other chatting sessions (e.g., ``[...] \textit{I appreciate you joining me. Today, I'll ask you some questions} [...]''). & 
 The chatbot's language emphasises the \textbf{persistent} nature of the conversation, and language alludes to conversational persistence (e.g., ``[...] \textit{I appreciate you joining me \textbf{again}. Today, I'll ask you some \textbf{more} questions} [...]''). \\
\bottomrule 
\end{tabular}
\end{table*}

\section{User Study}



In this study we investigate how both the framing of a chatbot's ephemerality (i.e., the (im)permanence of interactions, framed as either \textsc{Stranger} or \textsc{Familiar}) and the type of information disclosure (\textsc{Factual}- or \textsc{Emotional}-disclosure) influences user perceptions of self-disclosure. This gave us our research questions of:

\begin{itemize}
    \item \textbf{RQ1:} How does the ephemerality-framing of a chatbot (\textsc{Str\-anger}\slash\textsc{Familiar}) affect user perceptions related to self-dis\-closure?
    \item \textbf{RQ2:} How does the type of information disclosure (\textsc{Factual}\slash\textsc{Emotional}) to a chatbot affect user perceptions related to self-disclosure?
    \item \textbf{RQ3:} What is the combined effect of ephemerality-framing and disclosure type on user perceptions related to self-dis\-closure?
\end{itemize}


\begin{table*}[htbp]
\caption{Questions asked by the chatbot to elicit either \textsc{Emotional}- or \textsc{Factual}-disclosure.}
\label{tab:chatbot-questions}
\begin{tabular}{p{0.21\linewidth}p{0.76\linewidth}}
\toprule
 Disclosure Type & Chatbot Utterance \\
\midrule
 
 \multirow{4}{0.2\linewidth}{\textsc{Emotional}-Disclosure Questions} & 
 (1) ``\textit{How would you describe your current experience with feelings of loneliness or lack of social connection?}'' \\
 
 & 
 (2) ``\textit{What has your experience been like trying to form new, close relationships?}'' \\
 
 &
 (3) ``\textit{Is there anything holding you back from being as socially connected as you'd like to be?}'' \\
 
 &
 (4) ``\textit{Have you had any significant losses or broken relationships that still cause you pain or loneliness?}'' \\ \midrule
  
 \multirow{4}{0.2\linewidth}{\textsc{Factual}-Disclosure Questions} &
 (1) ``\textit{Can you name a few of your favourite books, movies, or TV shows?}'' \\
  
  &
 (2) ``\textit{Are you involved in any sports or physical activities?}'' \\
   
 &
 (3) ``\textit{What type of music do you enjoy listening to, and are there any artists you recommend?}'' \\
   
 &
 (4) ``\textit{Do you enjoy cooking or trying new recipes? What’s a dish you recently made?}'' \\
 
\bottomrule 
\end{tabular}
\end{table*}

Ethics approval was obtained from the institutional IRB. Although participants in the \textsc{Stranger} condition were told that the chatbot would not retain conversations, we logged all user utterances (in both \textsc{Stranger} and \textsc{Familiar} conditions) to support technical debugging and to compute word counts. The research team did not inspect utterance content.


\subsection{Experiment Conditions}

Participants on Prolific talked to a chatbot in two separate sessions across two days. 
We followed a 2~$\times$~2 mixed factorial design with a between-subjects variable of \textbf{Ephemerality-Framing} and within-subjects variable of \textbf{Disclosure Type}.  Experiment conditions are described below and illustrated in Figure~\ref{fig:Experiment-Flow}.

\subsubsection{Ephemerality-Framing:} 
\label{sec:Condition-Ephemerality}
Our between-subjects variable focuses on how the ephemerality of chatbot interactions are framed with users. 
\textbf{Ephemerality-Framing} has two levels of:
\begin{itemize}
    \item \textsc{Stranger}: The chatbot frames itself as a stranger to the user, with chatting sessions framed as independent from each other with no memory before or after the interaction.
    \item \textsc{Familiar}: The chatbot frames itself as familiar to the user, with chatting sessions framed as persistent and dependent on one another with memory of the user before and after the interaction.
\end{itemize}
See Table~\ref{tab:conversational-features} for differences in conversational features between the Ephemerality-framing conditions.


\begin{figure*}[htbp]
    \centering
    \begin{subfigure}[b]{0.49\textwidth}
        \includegraphics[width=\textwidth]{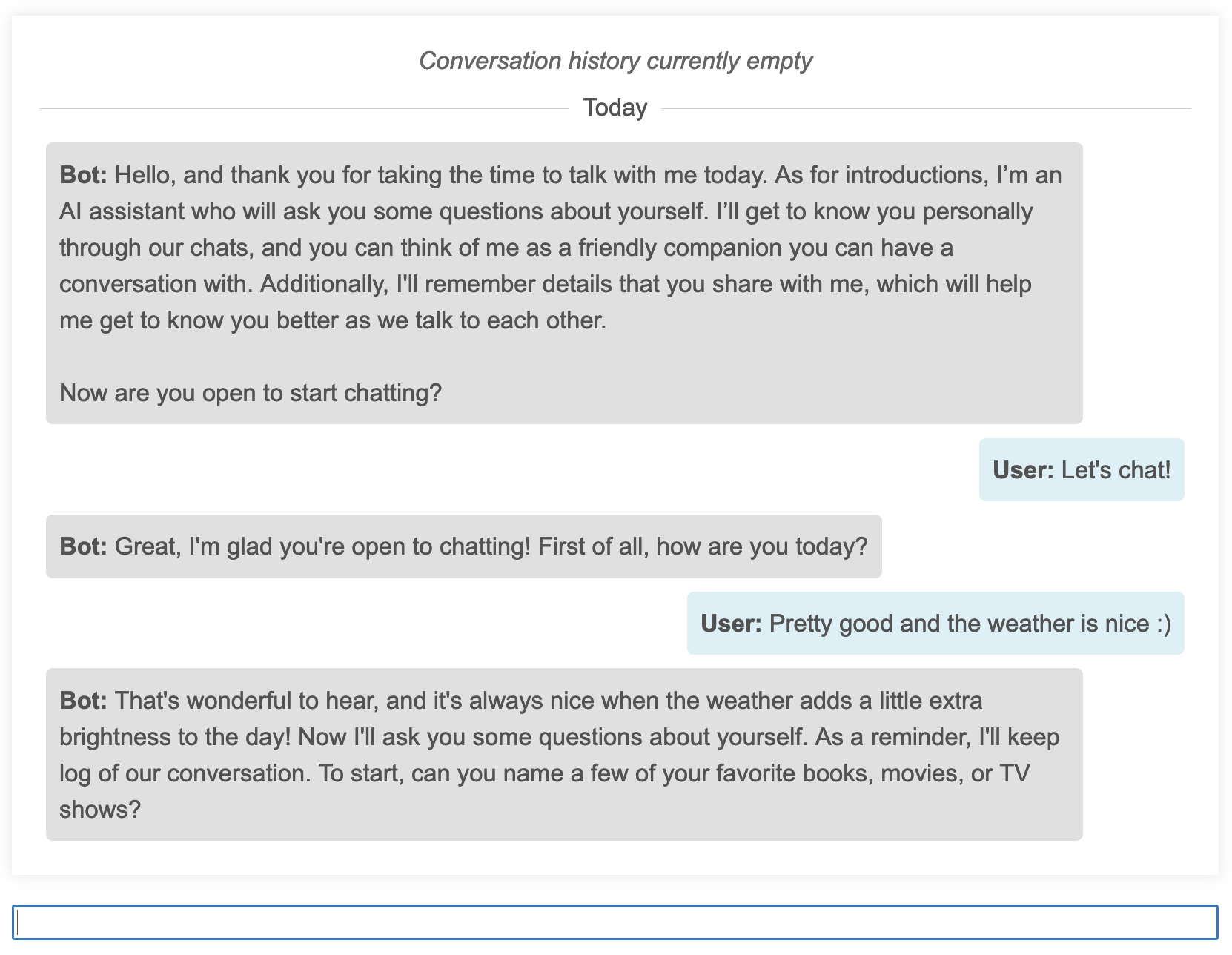}
        \caption{\textsc{Familiar} ephemerality-framing condition}
        \label{fig:Familiar-Session-1}
    \end{subfigure}
    \hfill 
    \begin{subfigure}[b]{0.49\textwidth}
        \includegraphics[width=\textwidth]{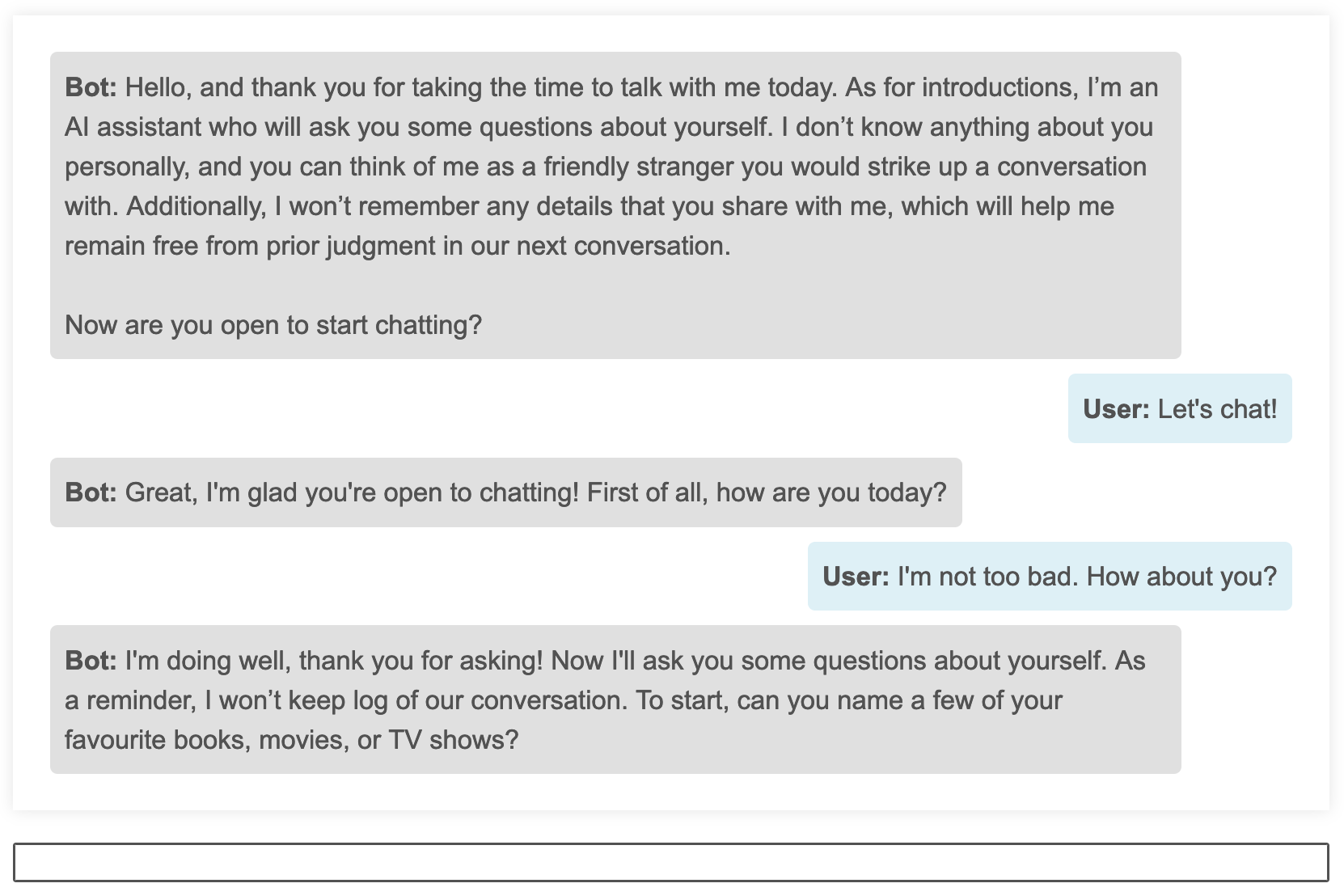}
        \caption{\textsc{Stranger} ephemerality-framing condition}
        \label{fig:Stranger-Session-1}
    \end{subfigure}
    \caption{The chatbot interface as seen by participants in the \textbf{first chatting session} for the \textsc{Factual}-disclosure condition. The chatbot initiates conversation by framing its relationship with ephemerality and the user.}
    \label{fig:interface-session-one}
\end{figure*}

\subsubsection{Disclosure Type:} 
\label{sec:Condition-Disclosure}
Our within-subjects variable controls the type of information requested during a chatting session. 
Namely, the chatbot will elicit either (1) low-sensitivity \textsc{Factual}-disclosure or (2) high-sensitivity \textsc{Emotional}-disclosure.
To motivate these conditions,
we followed prior work that found people perceived factual-disclosure (e.g., topics related to personal tastes such as preferences for food, music, or books) as less sensitive to self-disclose, and emotional-disclosure (e.g., topics related to interpersonal relationships and self-concept) as more sensitive to self-disclose~\cite{rubin1978friendship}.
Drawing on questions and topics from prior work~\cite{asher1984loneliness,ebesutani2012loneliness,rubin1978friendship}, \textsc{Factual}-disclosure consisted of questions related to hobbies and interests (e.g., ``\textit{What type of music do you enjoy listening to} [...]\textit{?}''), while \textsc{Emotional}-disclosure consisted of questions related to loneliness and social connections (e.g., ``\textit{What has your experience been like trying to form new, close relationships?}''). Questions asked per condition are shown in Table~\ref{tab:chatbot-questions}.

\subsection{Chatbot Scripting and Implementation}
\label{sec:scripting}


The chatbots were hosted on Qualtrics using HTML and JavaScript to emulate the look and feel of a chatbot (see Figure~\ref{fig:interface-session-one} for a comparison of interfaces between Ephemerality-Framing conditions). 
User responses were controlled so that only desktop or laptop devices could be used. For the \textsc{Familiar} condition, chatting history was retrieved automatically via participants' Prolific IDs.

Users talked with their assigned chatbot condition in two separate sessions across two days, with users being invited to the second session $\sim$24 hours after the first. 
To ensure experimental control and enable direct comparison between ephemerality conditions, the chatbot dictated the conversation flow. 
Additionally, questions asked to elicit self-disclosure (\textsc{Emotional} or \textsc{Factual}) were identical for all participants, consistent with prior work investigating self-disclosure~\cite{cox2022does,cox2023comparing,lee2020designing}. 
Chatbot scripting combined pre-scripted utterances and LLM-generated utterances (via GPT-4o~\cite{GPT-4o}).
Pre-scripted utterances established the chatbot's ephemerality-framing and elicited user self-disclosure, while LLM-generated utterances acknowledged user inputs.
See Tables~\ref{tab:script1} and~\ref{tab:script2} for scripts used in sessions one and two, respectively, alongside prompting used to generate GPT-4o utterances.

In two sessions ($\sim$24 hours apart), the chatbot elicited \textsc{Emotional}-disclosure in one session and \textsc{Factual}-disclosure in the other. 
Disclosure order was randomised, with half of participants experiencing \textsc{Emotional}-disclosure first and \textsc{Factual}-disclosure second, and the other half experiencing the reverse order.
Additionally to ease social interaction~\cite{bickmore1999small,lee2020hear}, each session began with small-talk (e.g., ``\textit{how are you today?}'') before seeking \textsc{Emotional}- or \textsc{Factual}-disclosure. 
This small-talk, intended for exchanging pleasantries, differs from \textsc{Factual}-disclosure questions, which gather specific personal details.

\subsection{Participants}

We recruited participants from Prolific using selection criteria to ensure reliable results (i.e., US-based\footnote{To control for cultural impact on self-disclosure, we chose only US-based participants. Potential implications and limitations of this are discussed in Section~\ref{sec:limitations}.},
English fluency, >97\% approval rate, >150 previous submissions). Participants were paid £1.20 to complete the first chatting session and £1.50 for the second, with average pay exceeding California minimum wage.
Excluding instruction and evaluation time, chatbot conversations lasted (median $\sim$3 min; mean $\sim$4 min), with no significant differences across conditions or sessions.
A total of 127 participants completed both sessions 
(mean age 36.1; 60 female, 65 male, 2 non-binary), and 9 participants completed only the first session. We include only data from participants who completed both sessions, resulting in 63 \textsc{Stranger} and 64 \textsc{Familiar} participants.

\begin{figure*}[htbp]
  \centering
  \includegraphics[width=0.6\linewidth]{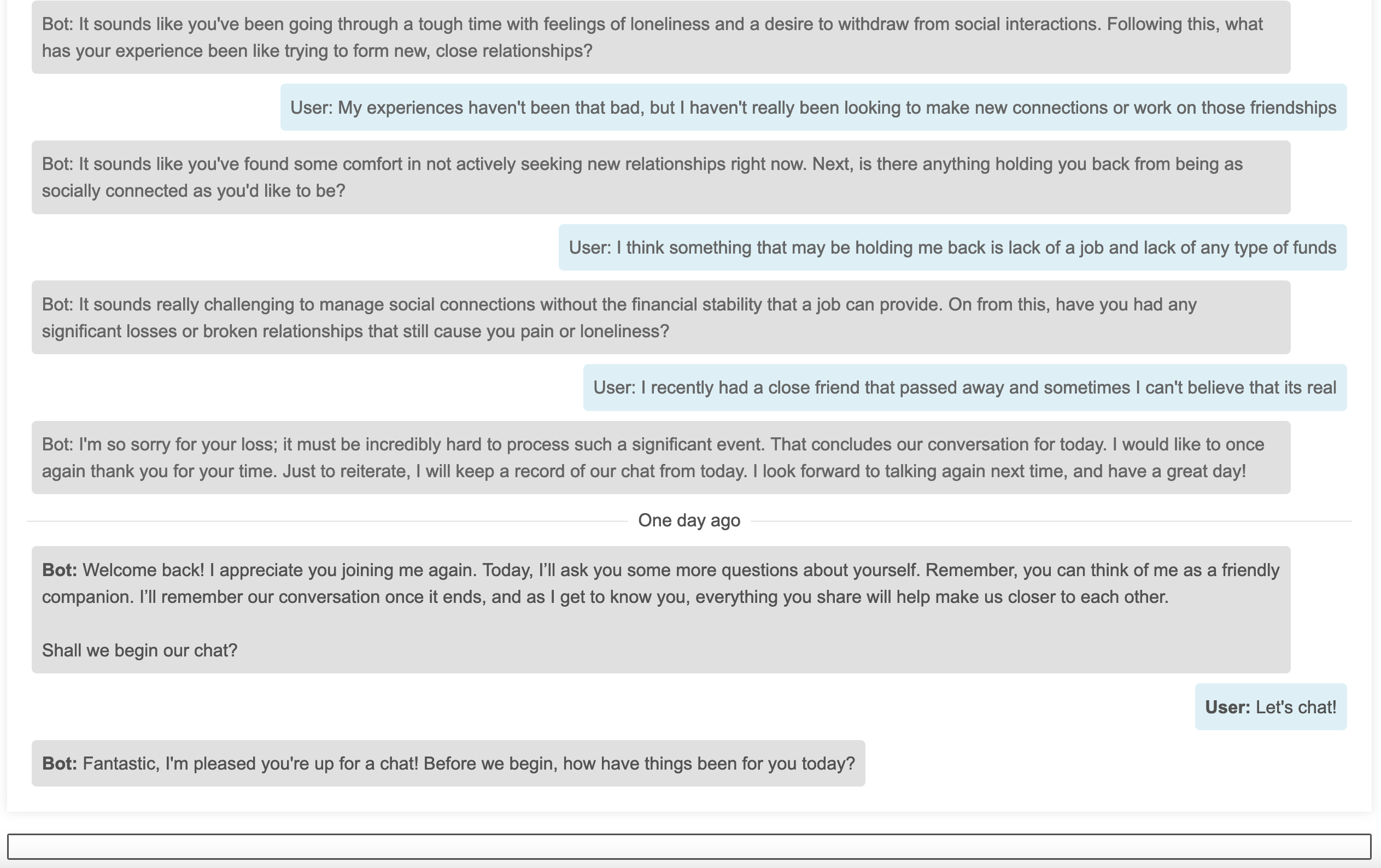}
  \caption{Chatbot interface as seen by \textsc{Familiar} participants in the \textbf{second session}. Chatting history (in this case from the \textsc{Emotional}-disclosure condition) is displayed above the current session's conversation.}
  \Description{Familiar Condition CA: A figure of the CA interface as seen by Familiar participants in the second chatting session. Chatting history is display above current conversation, and the CA initiates conversation by framing its relationship with ephemerality and the user.}
  \label{fig:interface-session-two-familiar}
\end{figure*}

\subsection{Procedure}

Participants followed the procedure below for both sessions:
\begin{enumerate}
    \item \textbf{Joining session:} Participant directed from Prolific to Qual\-trics (task named ``\textit{Talk to a chatbot about yourself (Part X of Two)}'' on Prolific). Participant receives high-level instructions.
    \item \textbf{Consent and demographics:} (\textit{Session one only}) participant completes consent form and demographics.
    \item \textbf{Task instructions:} Participant receives instructions (i.e., task description, reassurance of no right or wrong answers, and reminder that responses should be in English).
    \item \textbf{Chatbot interaction:} Participant talks with the chatbot from their Ephemerality-Framing condition.
    \item \textbf{Post-test questions:} Participant evaluates quantitative and qualitative experience (see Section~\ref{sec:measures}).
\end{enumerate}


\subsection{Measures}
\label{sec:measures}

We measured user perceptions of the chatbot and their relation to self-disclosure, along with open-ended feedback to explore \textit{why} these views were held. 
Additionally, response length was analysed using word count, which has served as an indicator of response quality in prior work~\cite{louw2011active,rhim2022application,barak2007degree}.

\subsubsection{Subjective Measures}


After each session, participants rated their perceived honesty, comfort in disclosing, closeness with the chatbot, enjoyment (a driver of continued interaction and self-disclosure~\cite{zhang2023dual,ledbetter2009measuring,lee2020hear,ho2018psychological}), and desire to continue using the chatbot.
Specifically, participants were asked ``\textit{To what extent do \textbf{you personally} agree or disagree that...}'' before rating the following on 7-point Likert scales (1 = Strongly Disagree, 7 = Strongly Agree):

\begin{itemize}
    \item \textbf{Honesty} of self-disclosure (4-items from Wheeless et al.~\cite{kang2024counseling,wheeless1976self,gibbs2006self}): ``\textit{I talked honestly about myself to the chatbot.}'', ``\textit{I was honest when I revealed my feelings and experiences to the chatbot.}'', ``\textit{I described myself accurately when talking about my feelings and experiences to the chatbot.}'', and ``\textit{The things I revealed to the chatbot are accurate reflections of who I really am.}'' ($M$ = 6.52, SD = 0.66, $\alpha$ = 0.916)\footnote{$\alpha$ denotes the Cronbach’s alpha reliability coefficient, which measures the internal consistency of a scale (i.e., the extent to which the items are correlated and measure the same construct). Values above 0.8 are generally considered good for empirical studies~\cite{lance2006sources}.}
    \item \textbf{Comfort} of self-disclosure (4-items~\cite{croes2021can,ledbetter2009measuring,croes202136,ischen2020privacy}): ``\textit{I felt like I could be personal during the interaction.}'', ``\textit{I felt comfortable disclosing personal information during the interaction.}'', ``\textit{It was easy to disclose personal information in the interaction.}'', and ``\textit{I felt like I could be open during the interaction.}'' ($M$ = 5.80, SD = 1.20, $\alpha$ = 0.924)
    \item \textbf{Enjoyment} talking to the chatbot (1-item~\cite{cox2022does,celino2020submitting}): ``\textit{I enjoyed talking to the chatbot}''
    \item \textbf{Desire to continue} using the chatbot (1-item~\cite{cox2022does,khadpe2020conceptual}): ``\textit{I would want to continue talking to the chatbot}'' 
\end{itemize}

Finally, to measure participants' \textbf{perceived closeness} to the chatbot, we used the inclusion of other in the self (IOS) scale, a validated pictorial scale~\cite{aron1992inclusion,gachter2015measuring}. 
Here, participants were asked to select the visual representation of two overlapping circles that best represented their relationship with the chatbot via the prompt ``\textit{Please select the picture below which best describes your relationship with the chatbot}''.

\subsubsection{Qualitative Measures}
\label{sec:QualMethods}

After collecting subjective ratings, participants provided open-text responses to offer further insights into their interaction with the chatbot.
Specifically, participants explained their IOS scale choice (``\textit{For your picture selection above, please explain your answer and what factors influenced your decision.}''); described what made them comfortable or uncomfortable (``\textit{What aspects of the conversation made you feel particularly \textbf{comfortable or uncomfortable sharing information}? Please elaborate on your answer.}''); and reflected on what influenced their willingness to self-disclose (``\textit{Please think back on your entire conversation with the chatbot. How did anything the chatbot said, including its introductions and responses, influence your willingness to share personal information?}'').

After the second (final) session, participants described changes in their experience between the two interactions, referencing the disclosure type condition 
(e.g., ``\textit{How did your experience with the chatbot change between your first and second interactions, if at all? What do you think influenced any changes? (For context, in the first session you discussed your personal interests, and in the second session you discussed social connections.)}'').

\begin{table*}[htbp]
\centering
\begin{tabular}{lllll}
    \toprule
     & \multicolumn{2}{c}{\textsc{Emotional}} & \multicolumn{2}{c}{\textsc{Factual}} \\
     \cmidrule(lr){2-3}\cmidrule(lr){4-5}
     & \textsc{Stranger} & \textsc{Familiar} & \textsc{Stranger} & \textsc{Familiar} \\
     \midrule
    Honesty & 6.61 (0.50) & 6.39 (0.69) & 6.56 (0.75) & 6.52 (0.64) \\
    Comfort & \textbf{5.98 (0.89)} & \textbf{5.54 (1.43)} & 5.90 (1.13) & 5.79 (1.27) \\
    Enjoyment & 4.94 (1.77) & 5.29 (1.69) & 5.05 (1.73) & 5.28 (1.60) \\
    Desire to Continue & 4.51 (1.90) &4.95 (1.83) & 4.71 (1.84) & 5.06 (1.81) \\
    Perceived Closeness & 2.51 (1.60) & 3.03 (1.66) & 2.54 (1.58) & 2.89 (1.76) \\
    Response Length & 86.3 (47.5) & 73.1 (43.9) & 80.9 (47.8) & 69.9 (44.0) \\
    \bottomrule
\end{tabular}
\caption{Outcome measures by \textit{Disclosure Type} and \textit{Ephemerality-Framing}. Values shown as ``Mean (SD)'', and significant differences ($p < 0.05$) shown in bold.}
\label{tab:summary-all}
\end{table*}

\begin{table*}[htbp]
\resizebox{\textwidth}{!}{ 
\centering
\footnotesize
\begin{tabular}{lllllllll}
    \toprule
    & \multicolumn{4}{c}{Session \textsc{one}} & \multicolumn{4}{c}{Session \textsc{two}} \\
    \cmidrule(lr){2-5}\cmidrule(lr){6-9}
    & \multicolumn{2}{c}{\textsc{Emotional}} & \multicolumn{2}{c}{\textsc{Factual}} & \multicolumn{2}{c}{\textsc{Emotional}} & \multicolumn{2}{c}{\textsc{Factual}} \\
    \cmidrule(lr){2-3}\cmidrule(lr){4-5}\cmidrule(lr){6-7}\cmidrule(lr){8-9}
    & \textsc{Stranger} & \textsc{Familiar} & \textsc{Stranger} & \textsc{Familiar} & \textsc{Stranger} & \textsc{Familiar} & \textsc{Stranger} & \textsc{Familiar} \\
    \cmidrule(lr){1-1}\cmidrule(lr){2-5}\cmidrule(lr){6-9}
    Honesty & 6.60 (0.57) & 6.34 (0.65) & 6.52 (0.68) & 6.58 (0.63) & 6.62 (0.42) & 6.43 (0.74) & 6.60 (0.82) & 6.44 (0.66) \\
    Comfort & \textbf{5.92 (0.86)} & \textbf{5.20 (1.51)} & 5.66 (1.22) & 5.88 (1.30) & 6.04 (0.94) & 5.87 (1.27) & 6.11 (1.01) & 5.69 (1.25) \\
    Enjoyment & 5.09 (1.79) & 4.88 (1.96) & 4.90 (1.58) & 5.55 (1.46) & \textbf{4.77 (1.76)} & \textbf{5.70 (1.29)} & 5.18 (1.86) & 5.00 (1.71) \\
    Desire to Continue & 4.73 (1.99) & 4.66 (1.94) & \textbf{4.43 (1.76)} & \textbf{5.48 (1.66)} & \textbf{4.27 (1.80)} & \textbf{5.24 (1.70)} & 4.97 (1.91) & 4.61 (1.87) \\
    Perceived Closeness & 2.15 (1.50) & 2.91 (1.65) & 2.60 (1.43) & 2.73 (1.68) & 2.90 (1.63) & 3.15 (1.68) & 2.48 (1.73) & 3.06 (1.84) \\
    Response Length & 83.0 (57.5) & 66.0 (44.5) & 81.9 (51.8) & 66.6 (35.9) & 89.9 (33.8) & 79.9 (42.8) & 80.0 (44.6) & 73.3 (51.6) \\
    \bottomrule
\end{tabular}
}
\caption{Outcome measures by \textit{Session Number}, \textit{Disclosure Type} and \textit{Ephemerality-Framing}. Values shown as ``Mean (SD)'', and significant differences ($p < 0.05$) shown in bold.}
\label{tab:summary-sessions}
\end{table*}

\subsubsection{Pre-Interaction Survey}

Following prior work investigating self-disclosure to chatbots~\cite{cox2022does}, we measured participants' belief in robotic intelligence and feelings via two items from Liu et al.~\cite{liu2018should}: ``\textit{I believe chatbots can be intelligent}'', ``\textit{I believe chatbots can have real feelings}'' on 7-point Likert scales.
To measure participants' tendency to disclose emotionally distressing experiences, we used the 12-item Distress Disclosure Index (DDI)\cite{kahn2001measuring,kahn2012distress}, that includes items such as: ``\textit{When I feel upset, I usually confide in my friends}'' and ``\textit{I prefer not to talk about my problems}''.
Here, a low DDI score indicates a tendency to not reveal emotional problems with others, while higher scores indicate a tendency to disclose.
Additionally, we recorded participant age and gender.

\subsubsection{Post-Interaction Survey}

As a manipulation check, the study's final question (end of session two) asked participants to specify the chatbot's ephemerality.
Specifically, participants were asked:
``\textit{Which of the following best describes the chatbot you interacted with?}'' with four possible answers of: 
``\textit{A) The chatbot was designed to remember our conversations and build on them over time. 
B) The chatbot treated each conversation as independent and did not retain information from previous interactions. 
C) I'm not sure about the chatbot's ability to remember our conversations. 
D) Other (please specify)}''.

\section{Results}

\subsection{Quantitative Findings}
To assess the impact of \textbf{Ephemerality-Framing} (\textsc{Stranger} \slash \textsc{Familiar}), \textbf{Disclosure Type} (\textsc{Factual}\slash\textsc{Emotional}), and their interaction across different chatting \textbf{Session Number}s (\textsc{one}\slash\textsc{two}) on the quantitative outcome measures, we used mixed-model ANOVA. 
This model was chosen due to the mixed design of our study, incorporating both fixed effects (Ephemerality-Framing, Disclosure Type, and Session Number) and random effects (Prolific ID nested within Ephemerality-Framing). This allowed us to account for the repeated measures within subjects and the variance attributable to individual differences.
We employed custom contrasts within a mixed-model ANOVA. This method was chosen due to its appropriateness for our factorial design, allowing for the control of Type I error across multiple comparisons.
This selective comparison allowed us to avoid dilution of effects through irrelevant comparisons.
Specifically, for our targeted contrasts we:
\begin{itemize}
    \item Compared Ephemerality-Framing conditions within the same Disclosure Type only (e.g., comparing differences in \textsc{Emotional}-disclosure between \textsc{Stranger} and \textsc{Familiar} conditions).
    \item Compared Session Numbers within the same Ephemerality-Framing, or within the same Disclosure Type (e.g., comparing differences between \textsc{Stranger} Session \textsc{one} and \textsc{Stranger} Session \textsc{two}).
\end{itemize}

\begin{figure*}[h]
  \centering
  \includegraphics[width=1\textwidth]{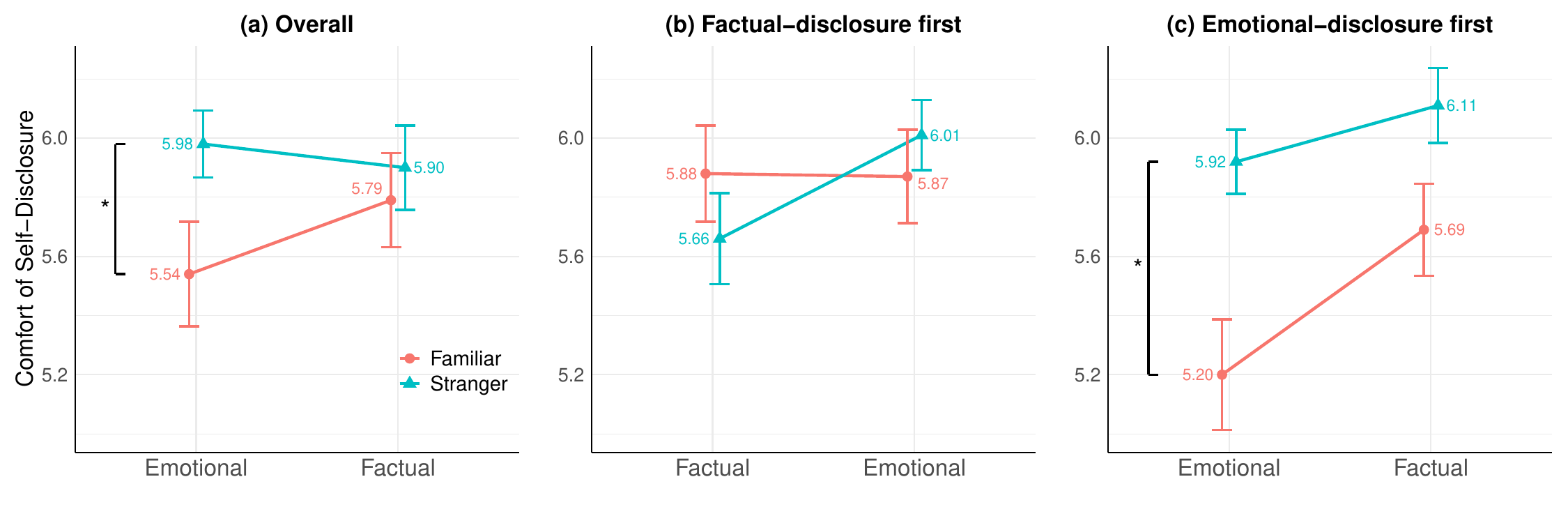}
  \caption{Plots of \textbf{Comfort of Self-Disclosure}. Error bars represent standard error, and significant differences ($p < 0.05$) indicated by square brackets. Participants felt more comfortable during \textsc{Emotional}-disclosure when talking to the \textsc{Stranger} chatbot. This effect was particularly pronounced when \textsc{Emotional}-disclosure was elicited in the first chatting session.}
  \Description{Plots of \textbf{comfort of self-disclosure}. Participants felt more comfortable during \textsc{Emotional}-disclosure when talking to the \textsc{Stranger} chatbot. This effect was particularly pronounced when \textsc{Emotional}-disclosure was elicited in the first chatting session.}
  \label{fig:comfort}
\end{figure*}

In the remainder of this section, $p$ represents the p-value, $M$ the mean, CI the confidence interval, and SD the standard deviation. Our analysis uses an alpha value of 0.05 for significance testing.
Please see Tables \ref{tab:summary-all} and \ref{tab:summary-sessions} for summary statistics by outcome variable.

\subsubsection{\textbf{Honesty} of Self-Disclosure}
There were no statistically significant differences between the Ephemerality-Framing conditions ($M_{Familiar}$ = 6.45, $M_{Stranger}$ = 6.58), or Disclosure Type conditions ($M_{Factual}$ = 6.50, $M_{Emotional}$ = 6.54). Similarly, there were no interaction effects between different conditions.

\subsubsection{\textbf{Comfort} of Self-Disclosure}

There was no statistically significant difference for Ephemerality-Framing ($M_{Familiar}$ = 5.65, $M_{Stranger}$ = 5.93, $p = .1387$) or for Disclosure Type ($M_{Factual}$ = 5.82, $M_{Emotional}$ = 5.76, $p = .4474$).
However, participants reported significantly higher levels of comfort during \textsc{Emotional}-disclosure to the \textsc{Stranger} chatbot as opposed to the \textsc{Familiar} chatbot ($M_{Familiar}$ = 5.53, $M_{Stranger}$ = 5.98, t(163.5) = -2.116, $p = .0358$, 95\% CI [-0.862, -0.030]). 
This effect suggests that the \textsc{Stranger} Ephemerality-Framing maintains comfort during \textsc{Emotional}-dis\-closure, whereas participants discussing \textsc{Emotional}-disclosures with the \textsc{Familiar} chatbot may feel reduced level of comfort.
Conversely, no significant difference was observed in comfort levels between the \textsc{Familiar} and \textsc{Stranger} conditions for \textsc{Factual}-disclosures ($p = .5474$), perhaps reflecting the less sensitive nature of \textsc{Factual}-disclosures~\cite{rubin1978friendship}.


However, once Session Number was taken into account, differences in perceived comfort between Ephemerality-Framing conditions was only found when \textsc{Emotional}-disclosure was sought in the first chatting session.
That is to say, participants were more comfortable discussing \textsc{Emotional}-disclosure with the \textsc{Stranger} chatbot compared to the \textsc{Familiar} chatbot when \textsc{Emotional}-disclosure was sought in session \textsc{one} ($M_{Familiar}$ = 5.20, $M_{Stranger}$ = 5.96, t(183.5) = -2.441, $p = .0157$, 95\% CI [-1.305, -0.138]).
In contrast, when \textsc{Emotional}-disclosure was sought in the second chatting session, there were no differences in comfort between the two Ephemerality-Framing conditions.
This relationship between Session Number and Disclosure Type are discussed in Section \ref{sec:discussion-ephemerality}. Please see Figure~\ref{fig:comfort} for plots of comforts, and Table \ref{tab:summary-sessions} for summary statistics by Session Number.
Additionally, there was a statistically significant difference between Session Numbers ($M_{one}$ = 5.66, $M_{two}$ = 5.92, $p = .0041$) with comfort increasing in session \textsc{two}. 
Within contrasting comparisons, this significance only held for \textsc{Stranger} ($M_{one}$ = 5.79, $M_{two}$ = 6.08, t(122.6) = -2.354, $p = .0202$, 95\% CI [-0.534, -0.046]).

\begin{figure*}[h]
  \centering
  \includegraphics[width=1\textwidth]{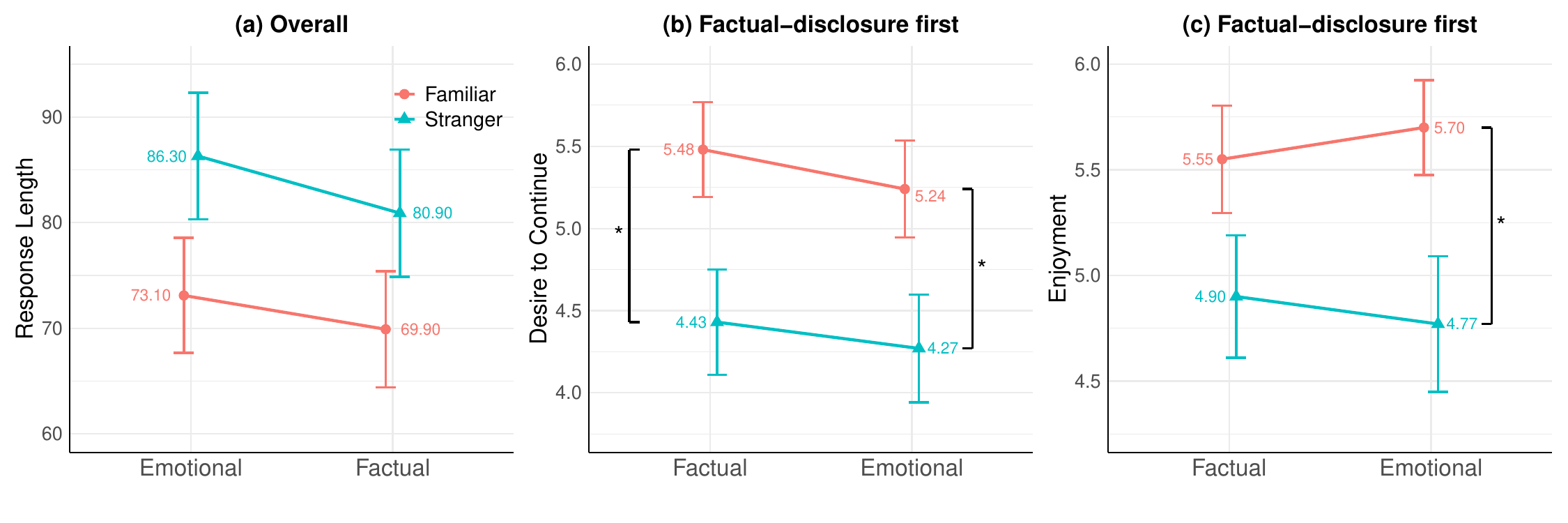}
  \caption{Plots of (a) \textbf{Response Length} (in both orderings), (b) \textbf{Desire to Continue} (\textsc{Factual} first), and (c) \textbf{Enjoyment} (\textsc{Factual} first). Significant differences ($p < 0.05$) indicated by square brackets. (b) and (c): \textsc{Familiar} received more positive engagement (desire to continue and enjoyment) than \textsc{Stranger} when ordering of disclosure type was \textsc{Factual}- followed by \textsc{Emotional}-disclosure, as would be seen in literature~\cite{altman1973socialpenetrationtheory}.}
  \Description{}
  \label{fig:quant_outcome_plots}
\end{figure*}

\subsubsection{\textbf{Enjoyment} talking to the chatbot}
There were no statistically significant differences in levels of enjoyment between Ephem\-erality-Framing conditions ($M_{Familiar}$ = 5.29, $M_{Stranger}$ = 4.98), or Disclosure Type conditions ($M_{Factual}$ = 5.16, $M_{Emotional}$ = 5.11). However, for chatting session \textsc{two}, when \textsc{Emotional}-disclosure was elicited there was a statistically significant difference between Ephemerality conditions ($M_{Familiar}$ = 5.70, $M_{Stranger}$ = 4.77, t(184) = 2.185, $p = .0303$, 95\% CI [0.090, 1.771]). Please see Figure \ref{fig:quant_outcome_plots}(c).
This effect suggests that when people are initially elicited for \textsc{Factual}-disclosure before being elicited for \textsc{Emotional}-disclosure, they will enjoy the chatbot more if it is framed as a \textsc{Familiar} interlocutor.
There was a significant positive association with belief in chatbot intelligence (Estimate = 0.397, SE = 0.084, t(121.1) = 4.71, $p < 0.0001$), suggesting that users who perceive chatbots as more intelligent enjoyed conversations more.


\subsubsection{\textbf{Desire to Continue} using the chatbot}
There were no statistically significant differences in desire to continue between Ephem\-erality-Framing conditions ($M_{Familiar}$ = 5.00, $M_{Stranger}$ = 4.60), or for Disclosure Type ($M_{Factual}$ = 4.88, $M_{Emotional}$ = 4.72).
Interestingly however, when users were elicited for \textsc{Factual}-disclosure in chatting session \textsc{one} before \textsc{Emotional}-disclosure in chatting session \textsc{two}, in \textit{both} chatting sessions participants had more desire to continue using the \textsc{Familiar} chatbot compared to \textsc{Stranger} (for \textit{chatting session \textsc{one}}: $M_{Familiar}$ = 5.48, $M_{Stranger}$ = 4.43, t(184) = 2.277, $p = .0241$, 95\% CI [0.139, 1.964]; and for \textit{chatting session \textsc{two}}: $M_{Familiar}$ = 5.24, $M_{Stranger}$ = 4.27, t(184) = 2.113, $p = .0362$, 95\% CI [0.064, 1.888]). 
Please see Figure \ref{fig:quant_outcome_plots}(b) for a plot of this relationship between Session Number and Disclosure Type.
There was a significant positive association with belief in chatbot intelligence (Estimate = 0.454, SE = 0.091, t(121.1) = 5.02, $p < 0.0001$), suggesting that users who perceive chatbots as more intelligent desired to continue usage.

\subsubsection{\textbf{Perceived Closeness} (IOS scale)}
Session Number had a statistically significant effect on perceived closeness, with participants rating increased closeness after session \textsc{two} compared to session \textsc{one} ($M_{one}$ = 2.60, $M_{two}$ = 2.89, $p = .003$, 95\% CI [0.104, 0.493]). Additionally (Session Number $*$ Ephemerality) was statistically significant for both ephemerality conditions, with session \textsc{two} scoring higher. 
There was a significant positive association with belief in chatbot intelligence (Estimate = 0.318, SE = 0.083, t(120.9) = 3.84, $p = 0.0002$), suggesting that users who perceive chatbots as more intelligent perceive themselves as closer to the chatbot.

\subsubsection{\textbf{Response Length} (word count)}

There were no statistically significant differences between Ephemerality-Framing conditions ($M_{Familiar}$ = 71.3, $M_{Stranger}$ = 83.7), or for Disclosure Type ($M_{Factual}$ = 75.2, $M_{Emotional}$ = 79.7), although trending differences match prior findings that personal questions elicit longer conversational responses~\cite{huang2023types}. 
Similarly, there were trending differences in response length between \textsc{Stranger} and \textsc{Familiar} when \textsc{Emotional}-disclosure was elicited ($M_{Familiar}$ = 73.1, $M_{Stranger}$ = 86.3, t(164) = -1.655, $p = .0995$, 95\% CI [-29.51, 2.58]).
There was a significant negative association with belief in chatbot intelligence (Estimate = -7.382, SE = 2.277, t(120.8) = -3.24, $p = 0.0015$), suggesting that users who perceive chatbots as more intelligent tended to use fewer words in their responses.




\subsection{Qualitative Findings}

As outlined in Section~\ref{sec:QualMethods}, we asked open-ended questions to explore users' perceptions of the chatbot, including what influenced their comfort, willingness to disclose, and sense of closeness with the chatbot.
To analyse open-ended responses, we followed an inductive thematic analysis approach~\cite{braun2006using}.
Specifically, two researchers independently familiarised themselves with all user responses, and generated initial codes and themes while remaining blind to the responses' experiment conditions. This was followed by discussions of theme interpretation and clarification with the research team.
When direct quotes are provided, we report the participant number followed by the chatting session, ephemerality-framing, and disclosure type in brackets.
For example, ``P10(\textsc{S2-Stranger-Factual})'' refers to participant 10 in session \textsc{two} under the \textsc{Stranger} and \textsc{Factual} conditions.


\subsubsection{\textbf{Information Disclosure in Conversations}}

Participants described how the type of information disclosure affected factors such as their feelings of connection with the chatbot, comfort, and willingness to self-disclose. 
Consistent with prior work~\cite{rubin1978friendship}, \textsc{Factual} disclosures were seen as low sensitivity and \textsc{Emotional} as high, with slight variations based on the participant's Ephemerality-Framing condition. We elaborate on this below.

Participants in both Ephemerality-Framing conditions found \textsc{Factual}-disclosure easy and comfortable to share, citing the non-intrusive questions as fostering comfort and reducing judgement or privacy concerns.
For example, P54(\textsc{S1-Familiar-Factual}) stated: ``\textit{I felt comfortable sharing because the questions were easy and not too personal. I mean, what risk could come from the world reading what book I like?}''.
Similarly, P81(\textsc{S1-Stranger-Factual}) noted: ``\textit{Asking of my favourite food made me comfortable and relaxed} [...]''.
However, what some participants perceived as non-intrusive, others found unengaging and ``\textit{shallow}'', with some describing feeling ``\textit{bored}'' or ``\textit{less interested}'' in the conversation. 
As P18(\textsc{S2-Stranger-Factual}) explained, the \textsc{Factual} session was: ``\textit{more boring because the questions were less introspective}''.
Additionally, some participants described feeling less perceived closeness with the chatbot when \textsc{Factual}-disclosure was sought. For example, P91(\textsc{S2-Stranger-Factual}) stated: ``\textit{I definitely felt separate from the bot today, as the} [\textsc{Factual}-disclosure] \textit{questions it asked felt more disconnected} [compared to \textsc{Emotional}-disclosure]''.

In contrast, \textsc{Emotional}-disclosure elicited more varied responses between Ephemerality-Framing conditions, with concerns surrounding sensitivity and comfort being more pronounced in the \textsc{Familiar} condition.
As P21(\textsc{S2-Familiar-Factual}) noted:
``\textit{The first} [\textsc{Emotional}] \textit{conversation felt more intrusive whereas I didn't mind sharing my interests in the second} [\textsc{Factual}] \textit{conversation}''.
Further, some \textsc{Familiar} participants were less forthcoming with \textsc{Emotional}-disclosure, 
such as P88(\textsc{S2-Familiar-Emotional}): ``\textit{I was hesitant to elaborate on details of my personal relationships, so I kept some responses vague}'', and P125(\textsc{S2-Familiar-Emotional}) who described not disclosing ``\textit{names or saying anything identifiable}''.
In addition, feelings of intrusiveness among (\textsc{Familiar-Emotional}) participants was affected by the order of disclosure sought by the chatbot.
Specifically, \textsc{Familiar} participants preferred \textsc{Factual}-disclosure being sought first, with several describing this as helping relationship development:

\begin{displayquote}
    ``\textit{I was more willing to open up after having the initial conversation. Starting off with `safe/easy' topics such as books and music allowed me to become familiar} [...]. \textit{I might have not been as open had the initial conversation been about social connections.} [...]'' - P54(\textsc{S2-Familiar-Emotional})
\end{displayquote}

Conversely, when \textsc{Emotional}-disclosure was sought first, \textsc{Familiar} participants described feelings of intrusiveness, as illustrated by P5(\textsc{S1-Familiar-Emotional}): ``\textit{I think if it started out by asking not as personal questions I would have been more willing} [to disclose]''.

This contrasts with \textsc{Stranger} participants who sometimes described no difference in comfort between disclosure-types, 
or described feeling more comfortable during \textsc{Emotional}-disclosure
(e.g., P83(\textsc{S2-Stranger-Factual}): ``\textit{The first} [Emotional] \textit{conversation was a lot more personable and made me feel more comfortable.}''). 
Further, several participants preferred \textsc{Emotional}-disclosure, describing it as ``\textit{personable}'', ``\textit{deep}'', ``\textit{genuine}'' and stimulating compared to \textsc{Factual}-disclosure.
For some, this fostered a stronger sense of closeness to the chatbot, such as P35(\textsc{S2-Stranger-Emo\-tional}): 
``\textit{the conversation felt personal, and really dug deep into certain emotions, creating a strong sense of friendship}''.
Finally, P12(\textsc{S2-Stranger-Emotional}) found the ``\textit{impartial}'' nature of the \textsc{Stran\-ger} conversation comforting when sharing \textsc{Emotional}-disclosure: 
``\textit{This conversation went more in depth about feelings. It was easier to share information with an impartial listener}''.


\subsubsection{\textbf{Chatbot Memory, Data Retention, and Continuity of Interactions}}

Participants described how the chatbot's ephem\-erality-framing, memory between sessions (e.g., how conversations are stored or shared), and repeated interactions affected feelings such as comfort and willingness to disclose, honesty, and closeness.

\textsc{Stranger} participants often described feeling comfortable during conversations, 
and frequently cited the lack of data retention or sharing as a key factor in their willingness to disclose, for instance: 
\begin{displayquote}
``\textit{It wouldn't remember anything so I YOLO'd and shared personal info}'' - P98(\textsc{S1-Stranger-Emotional})
\end{displayquote}
On from this, \textsc{Stranger} participants described a variety of reasons such as having an increased sense of anonymity, and reduced fear of social judgement or negative privacy consequences.
For example, P36(\textsc{S2-Stranger-Emotional}) described: ``\textit{I liked knowing that things I said to the chatbot would not become subjects gossip among common acquaintances}''.
Additionally, a common sentiment among \textsc{Stranger} participants was a lack of judgement from the chatbot, as described by P124(\textsc{S2-Stranger-Emotional}): ``\textit{I felt comfortable knowing I would not be judged and that it would not be remembered. With that it was easier to open up with knowing I did not have to worry about anything later}''.
Similarly, some \textsc{Stranger} participants stated that the lack of data retention made them more honest in their responses, such as P103(\textsc{S1-Stranger-Emotional}): ``\textit{The fact that my thoughts were not collected and saved made me okay with being honest with my answers}''.


Furthermore, multiple participants described the \textsc{Stranger} condition as comfortable to disclose to, due to the similarity in feeling to ``\textit{journaling}'' or ``\textit{diary}'' keeping, such as:
\begin{displayquote}
    ``\textit{I didn't feel judged} [...] \textit{I felt like I was just journaling with helpful prompts being offered along the way, and that made me feel more comfortable sharing information.}'' - P10(\textsc{S1-Stranger-Emotional})
\end{displayquote}
Interestingly, \textsc{Stranger} also had the effect of priming some participants to expect a sensitive conversation (due to the language and framing of the chatbot not remembering user utterances). This then led to a negative expectancy violation~\cite{burgoon2016application} when the first chatting session instead asked for factual-disclosure, as described by P30(\textsc{S1-Stranger-Factual}): ``\textit{It mentioned it wouldn't save the conversation, and then all its questions were very broad and nothing that I would consider a secret}''.

In the \textsc{Familiar} condition, participants expressed mixed feelings. While some appreciated the chatbot’s continuity between conversations, familiarity and perceived friendliness, others felt uneasy knowing that their responses were retained.
Additionally, some participants indicated a lack of trust in \textsc{Familiar} despite the tone being seen as friendly. For instance P14(\textsc{S2-Familiar-Factual}) stated:
``\textit{i think it's getting friendly but it's just harvesting some info}''.
Concerns about privacy, data usage, and potential misuse were more prominent in \textsc{Familiar}, especially when discussing emotional topics. For example, P82(\textsc{S1-Familiar-Emotional}) described that they felt:
``[...] \textit{uncomfortable knowing its a bot and it can disclose my information}''.

In both ephemerality-framing conditions, participants described how they felt closer to the chatbot after the second chatting session due to developing some level of familiarity during repeated interactions (mirroring findings from prior work~\cite{croes2021can}).
This was particularly pronounced for \textsc{Familiar} participants, where multiple participants described feeling closer due to the chatbot remembering conversations (e.g., P80(\textsc{S2-Familiar-Factual}): ``\textit{The fact that it continued our previous conversation was a big part of it}''), as well as the framing of the relationship (e.g., P116(\textsc{S2-Familiar-Factual}): ``\textit{I feel closer to the chatbot after it framed our chats as developing a relationship}'').
In contrast, P43(\textsc{S2-Stranger-Factual}) described \textsc{Stranger}'s lack of memory as hampering feelings of closeness:
``\textit{It still couldn't be a huge relationship blending because it's not a relationship - it doesn't even remember what you said previously. While in a way that is refreshing, it isn't relationship building}''.



\subsubsection{\textbf{User Beliefs in Chatbots}}

Participants described how their belief in chatbots possessing feelings influenced their perceived closeness, and their comfort and willingness to disclose.
Several participants noted their disbelief in chatbot emotions reduced their sense of closeness, citing the chatbot's lack of ``\textit{authenticity}''.
For example, P63(\textsc{S2-Familiar-Emotional}) stated: 
``\textit{The chatbot is a good listener, which makes me feel some closeness. However, I know that its feelings and empathy are not real, so I do not feel especially close}''. 
This also ranged to some participants describing how a perceived lack of lived experiences or tacit knowledge on the part of the chatbot lessened their feeling of closeness, with P89(\textsc{S1-Familiar-Factual}) describing: ``\textit{It's a chatbot, not a person.} [...] \textit{it can't have its own feelings on a topic which would let us continue discussing something. I can't spontaneously learn it likes X or dislikes Y, because it actually can't have those feelings} [...]''.
This is in contrast to some participants who described feeling less close to the chatbot due to its lack of reciprocal self-disclosure.
Although there is evidence that reciprocal disclosure from chatbots can increase user disclosure~\cite{moon2000intimate,lee2020hear}, this highlights the nuance in relation to users' beliefs regarding chatbot abilities. This also mirrors prior discussion that social exchange theories are not applicable to human-chatbot relationships, as chatbots do not experience costs or rewards from self-disclosure~\cite{fox2021relationship,skjuve2022longitudinal}.

In contrast to these negative perceptions, some participants described how the chatbot’s non-human nature made them feel more at ease. 
For example, P107(\textsc{S1-Familiar-Factual}) described not feeling judged as their interlocutor is not human: 
``\textit{I was comfortable sharing information because I knew that the chatbot wasn't a real person. It doesn't have real feels and can't really judge me}''.
Beyond this, some participants described lessened feelings of social risks in talking to a chatbot compared to humans, as well as a lack of guilt in burdening a human interlocutor.
For example, P55(\textsc{S1-Stranger-Emotional}) stated: 
``\textit{It was nice to talk to someone who I didn't think had too much of its own problems. I didn't feel like a burden} [...] \textit{I worry too much about someone using what I say against me later}''.
Additionally, some participants described the greater comfort and ease in self-disclosing sensitive personal information to a chatbot rather than a human, similar to findings of prior work~\cite{pickard2016revealing,lucas2014s,bowman2024exploring,jacobsen2025chatbots}.

\section{Discussion}


This work explores how a chatbot's ephemerality-framing influences self-disclosure.
By \textit{framing}, we mean how a chatbot \textit{describes itself conversationally}, shaping the relationship users expect~\cite{khadpe2020conceptual}.
Participants interacted with either a \textsc{Stranger} chatbot (framed as a temporary partner that forgets interactions) or a \textsc{Familiar} chatbot (framed as an enduring companion that remembers).
Across two sessions, the chatbot elicited either \textsc{Emotional}- or \textsc{Factual}-disclosure.
Our findings challenge the commonly held assumption that chatbots should always aim for greater personalisation and familiarity, showing that while familiarity can be beneficial, it may not consistently enhance self-disclosure depending on the context.

\subsection{Summary of Findings}


The key findings in relation to research questions one to three are as follows:

\noindent\textbf{RQ1: How does ephemerality-framing (\textsc{Stranger}\slash\textsc{Familiar}) affect perceptions related to self-disclosure?}
\begin{itemize}
    \item \textbf{Quantitatively,} participants were more comfortable sharing \textsc{Emotional}-disclosures under the \textsc{Stranger} framing, but only when \textsc{Emotional}-disclosure was sought first.
    \item \textbf{Qualitatively,} \textsc{Stranger} felt anonymous and judgement-free, lowering perceived social risks. 
    \textsc{Familiar} was seen as friendly by some, but also felt intrusive and raised privacy concerns (especially when \textsc{Emotional}-disclosure came first).
\end{itemize}

\noindent\textbf{RQ2: How does disclosure-type (\textsc{Factual}\slash\textsc{Emotional}) affect perceptions related to self-disclosure?}
\begin{itemize}
    \item \textbf{Quantitatively,} there were no \textit{overall} differences in measures between \textsc{Factual}- or \textsc{Emotional}-disclosure. Rather, comfort was moderated by ephemerality-framing and question order (see RQ1 and RQ3).
    \item \textbf{Qualitatively,} \textsc{Factual}-disclosure was seen as low-risk and ``\textit{easy}'', but sometimes shallow. \textsc{Emotional}-disclosure felt more engaging, but feelings of comfort were moderated by ephemerality-framing (see RQ1 and RQ3).
\end{itemize}

\noindent\textbf{RQ3: What is the combined effect of ephemerality-framing and disclosure type on user perceptions related to self-dis\-closure?}
\begin{itemize}
    \item \textbf{Quantitatively,} when \textsc{Emotional}-disclosure was sought first, users were more comfortable under the \textsc{Stranger} framing. However, when \textsc{Factual}-disclosure was sought first, the comfort gap closed and the \textsc{Familiar} framing yielded higher enjoyment and willingness to continue. Across both orders, perceived closeness grew in the second session.
    \item \textbf{Qualitative} findings matched quantitative. I.e., starting with low-risk \textsc{Factual}-disclosure let users establish initial rapport with the \textsc{Familiar} chatbot, leading participants to feel more at ease when \textsc{Emotional}-disclosure was later sought.
\end{itemize}

In addition, participants' \textbf{belief in chatbot emotions and intelligence} impacted perceptions. Quantitatively, participants who believed that chatbots are more intelligent: enjoyed conversations more, had stronger desire to continue, had greater perceived closeness, and provided shorter responses to the chatbot.
While shorter responses may seem counter-intuitive, this mirrors prior work showing that people who trust their interlocutor’s intelligence (whether human~\cite{clark_grounding_1991} or chatbot~\cite{cox2023comparing}) omit peripheral details, trusting them to fill in missing context.
Qualitatively, belief in the chatbot’s emotional capacity cut two ways. Some participants said that perceiving the chatbot as capable of real feelings increased their willingness to disclose, whereas others preferred interacting with a clearly non-human agent, citing reduced social risk and judgement when emotions felt obviously simulated.



\subsection{Ephemerality in Chatbot Conversations}
\label{sec:discussion-ephemerality}
Our findings align with the ``\textit{stranger on the train}''~\cite{altman1973socialpenetrationtheory,perlman1987revelation,derlega1977privacy} and ``\textit{stranger on the Internet}''~\cite{duradoni2021stranger,misoch2015stranger} phenomena, 
where people often feel comfortable disclosing to an ephemeral interlocutor.
Specifically, our quantitative results show that when \textsc{Emotional}-disclosure was sought in the first chatbot interaction, participants felt more comfortable sharing with the \textsc{Stranger} chatbot than with the \textsc{Familiar} chatbot.
This reflects transient, low-risk encounters where the lack of long-term consequences reduces social risks and encourages openness~\cite{misoch2015stranger,derlega1977privacy}.
Qualitatively, participants in the \textsc{Stranger} condition reported feeling anonymous, less judged, and freer to disclose, knowing their disclosures would not persist. 
This mirrors human interactions, where fleeting connections lower the perceived risks of vulnerability~\cite{ma2019effects,altman1973socialpenetrationtheory}.
Our results suggest this effect extends to human-chatbot interactions, where ephemeral framing may more effectively promote disclosure in sensitive contexts, such as emotional disclosure, mental health support, or journalling~\cite{lee2020designing,rashik2025ai}.
Within journalling contexts, not keeping a record of conversations may seem to contradict recent CA work~\cite{rashik2025ai}.
Yet, as in ``\textit{expressive writing}'' paradigms~\cite{Baikie_Wilhelm_2005,park2021wrote}, merely articulating thoughts and feelings yields emotional\footnote{Cf. Park et al.’s CA study~\cite{park2021wrote}, which found that expressive writing prompts \textit{without} responsive follow-ups led to greater emotional disclosure and less writing difficulty.} and physical benefits, independent of later re-reading. 
Permanent records then offer additive benefits (memory preservation, and longitudinal reflection), but these sit atop the therapeutic benefits of uninhibited expression.
Therefore, while personalisation can build rapport, framing a chatbot as a `\textit{stranger}' may offer a safer, more cathartic space for open sharing, free from concerns of future repercussions (as observed in social networking site (SNS) ephemerality~\cite{bayer2016sharing,yao2024s}).

On from this, our findings align with Social Penetration Theory ~\cite{altman1973socialpenetrationtheory}, which suggests (as discussed in Section \ref{sec:RW_self-disclosure}) that self-dis\-closure is influenced by the stage and intimacy of a relationship~\cite{sprecher2004self}, as well as perceived social costs and benefits.
That is to say, the order that either \textsc{Emotional}- or \textsc{Factual}-disclosure was sought affected participants' feelings of comfort in disclosing. 
As described above, if the \textit{first interaction} sought \textsc{Emotional}-disclosure, participants in the \textsc{Stranger} condition felt more comfortable compared to participants in the \textsc{Familiar} condition (where participants described less willingness to disclose due to feelings of intrusiveness).
However, if \textsc{Factual}-disclosure was sought in the first chatting session followed by \textsc{Emotional}-disclosure in the second chatting session, these differences in comfort between the \textsc{Stranger} and \textsc{Familiar} conditions were no longer significant. 
Participants in the \textsc{Familiar} condition described this order of disclosure as being appropriate for a developing relationship (as may be expected from both human-human~\cite{altman1973socialpenetrationtheory} and human-chatbot relationship building~\cite{skjuve2021my}), in addition to feeling more enjoyment and desire to continue.
This also coincided with quantitative and qualitative findings that participants across conditions felt closer to the chatbot in their second chatting session (similarly to prior findings that people feel closer after repeated interactions~\cite{croes2021can}).
Our qualitative findings also aligned with previous research, showing that discussing sensitive topics fosters a greater sense of closeness to a conversational agent~\cite{cho2019hey}.
These findings suggest that the initial context of the interaction plays a critical role in shaping user comfort when self-disclosing. 
Once users had been familiarised with the chatbot through prior \textsc{Factual}-disclosures, the perceived distinction between an ephemeral and persistent interaction diminished. 
This finding suggests that while ephemeral-framing may be more effective for encouraging emotional-disclosure in first-time interactions, the need for such framing may decrease as users become more accustomed to the chatbot over time. 
Designers may therefore consider using ephemeral framing selectively, particularly in initial or one-off encounters, while allowing the chatbot to gradually build familiarity in subsequent interactions.




Additionally, the chatbot’s \textit{act of saying} that utterances would not be remembered seemed to implicitly prime some users for more sensitive conversations. This suggests that merely informing users about the ephemerality of the interaction led to a tacit assumption that sensitive topics were to follow. The very acknowledgement of memory retention or lack thereof may have created expectations regarding the depth of the conversation, further influencing how users approached the interaction. Similar to prior work~\cite{khadpe2020conceptual,hartmann2008framing}, this priming effect highlights the importance of carefully managing the way a chatbot presents its capabilities to avoid inadvertently shaping user expectations in unintended ways.

Finally, user comfort in emotional-disclosure to the ephemeral chatbot raises discussion as to whether sensitive conversations should be stored or utilised by LLM providers.
Relatedly, it has been shown that atypical user utterances can be regurgitated by language models, which can then be exploited by adversarial parties~\cite{carlini2021extracting}. 
This has led some researchers to call for only publicly available text content to be used to train models to avoid unintended leaks of personal and compromising information~\cite{brown2022does}.
These concerns highlight the need to reconsider storing all user conversations and emphasise the importance of privacy-focused data retention policies, along with clear assurances on data processing by LLM providers.

\subsection{Future Opportunities}
\label{sec:discussion-future}

In terms of memory between chatting sessions, our study followed a somewhat similar ephemerality to current LLMs (such as ChatGPT's ``\textit{temporary chat}''~\cite{ChatGPT-memory-article} being somewhat equivalent to \textsc{Stranger}). Here, ephemerality and memory are treated in a more binary sense with either presence or absence of memory in chatting sessions.
Future studies could explore adaptive ephemerality based on conversational context, allowing automated adjustments and user control over data retention (e.g., options to forget sensitive conversations). This approach could mirror the effectiveness of ephemeral sharing in SNSs, which enhances engagement and disclosure~\cite{chiu2021last,xu2016automatic}.
Additionally, such user control would match calls for empowerment for people to control their own data~\cite{seberger2021empowering}.
Future opportunities could take influence from prior work within SNSs, where there has been discussion as to whether our personal data should ``\textit{decay}'' over time~\cite{mohamed2020influence,yamakami2010we}. 
For example, as conversational history ages, only the most significant (while non-sensitive) conversational features may want to be retained in order to allow for personalisation while still retaining a sense of privacy.
A decaying form of memory would also match the desire of SNS users who have described wanting their online presence to decay in line with their view of human memories (similarly to human memory decay theory)~\cite{mohamed2018online}.
Data retention and ephemerality could also follow a more flexible natural language approach, such as chatbots explicitly asking users if conversations should be remembered (e.g., ``\textit{Would you like me to keep this off the record?}''), or allowing for users to ad hoc request that conversations not be stored (e.g., ``\textit{Please keep this conversation off the record.}'').
However, increasing users' sense of control over their data may paradoxically create privacy risks by fostering overconfidence in data protection~\cite{brandimarte2013misplaced}.

Additionally, future studies could explore different cues~\cite{doring2013ephemeral}, metaphors~\cite{jung2022great,desai2023metaphors}, or personas~\cite{kollerup2025enhancing} to signal ephemerality to users.
For example, visual cues~\cite{doring2013ephemeral} could be used in embodied agents to signal ephemerality (such as avoiding eye-contact with the user which has been found to increase self-disclosure~\cite{yuan2024don}).
Future studies could construct the framing of ephemerality more precisely such as specifying which data would be remembered when talking to the chatbot, or how memories would be stored and referenced~\cite{cox2023comparing}.
The metaphorical ``setting'' of disclosure could be explored. While in-person settings will be recognisable as more ephemeral or anonymous (such as anonymous hotlines or group therapy sessions where anonymity fosters self-disclosure) similar settings may be less explored within the context of chatbots. Relatedly, recent work has investigated people's self-disclosure (or ``\textit{confessions}'') to a ``\textit{priestess}'' chatbot~\cite{croes2024digital}, yet comparative forms of metaphor and their associated anonymity and ephemerality have been less explored.


Further, additional research is needed to understand how self-disclosure dynamics evolve over multiple sessions.
In the case of this study, if only one chatting session were used, the \textsc{Stranger} condition would have been reported as more comfortable. However, our experimental design highlights the nuanced effects of multiple interactions with a conversational agent.
On from this, while relationship development typically follows a progression from more surface level to intimate disclosures~\cite{altman1973socialpenetrationtheory,skjuve2021my}, over time self-disclosure in established relationships may exhibit a cyclical pattern~\cite{sprecher2004self}. For instance, there may be phases where one partner seeks privacy or self-reflection, resulting in a temporary reduction in disclosure. These phases may later give way to renewed openness and deeper emotional sharing, driven by changing contexts or relational needs.
This highlights the complexity and nuance inherent in relationships over multiple interactions. 
Future research could investigate how evolving personal circumstances and relationship dynamics influence patterns of self-disclosure.
Additionally, future work could explore how factors (such as ephemerality) influence these cycles of disclosure within relationships. 
For example, does framing an interaction as ephemeral or persistent amplify or dampen the natural oscillation between phases of openness and privacy? 
Such studies would provide deeper insights into how design features can support or inhibit the dynamic nature of disclosure over time.

\subsection{Limitations}
\label{sec:limitations}

We highlight several limitations in our study.
First, participants did not interact with the study's chatbot as part of a real-world deployment. 
\textcolor{edit}{Instead, participants talked to a chatbot embedded in Qualtrics as part of a controlled academic study. Here, both the setting of the interactions (i.e., via Qualtrics), and the institution hosting the chatbot~\cite{asbjorn2018trust,10.1145/3708359.3712125} could affect people's trust and behaviour.}
Beyond this, there is evidence of people already sharing sensitive information in conversations with LLMs~\cite{zhang2024s,ngongprotecting} thereby drawing attention to the potentially complicated nature of self-disclosure given human-like and believable chatbot interlocutors.
Additionally, the lack of reciprocal self-disclosure~\cite{moon2000intimate,lee2020hear,meng2021emotional} from the \textsc{Familiar} chatbot (due to the controlled nature of our experiment) may have been unexpected to participants who anticipated a familiar interlocutor to follow such social norms~\cite{croes2021can}.

Furthermore, all participants were US-based, meaning results may not generalise to cultures with different attitudes towards chatbots~\cite{liu2024understanding,zhang2024ai} or self-disclosure~\cite{schug2010relational,zhao2012and,de2017gender,kito2005self,tsai2018culture}.
Specifically, prior work has found cultures may vary in both their extent of emotional disclosure~\cite{tsai2018culture}, and to what extent people are likely to self-disclose to various people (such as strangers, colleagues, or close family)~\cite{zhao2012and,kito2005self}.
Additionally, as the US is a culture with high relational mobility (where relationships can be formed and dissolved relatively easily) people may disclose more to foster commitment and strengthen their relationships~\cite{schug2010relational}.

Similarly to prior work~\cite{cox2023comparing,chen2024different}, we did not measure self-disclosure quality, but rather users' \textit{perceptions of} and \textit{beliefs towards} self-disclosure. This is as we want to investigate the user's feelings about the natures of their disclosures, regardless of the exact content of disclosure. From this we take both a qualitative and focused subjective way of measuring the effect of ephemerality-framing.
Finally, there was no baseline condition that was free of ephemerality-framing (e.g., a chatbot simply stating that it would talk to users). 
However, such a baseline could be influenced by one's preconceived notions of ephemerality (such as whether a chatbot is seen as a companion that remembers interactions). To control for this variability, we employed distinct ephemerality-framing conditions.

\section{Conclusion}

This study investigates the impact of a chatbot's ephemerality-framing on self-disclosure.
By ephemerality-framing, we refer to \textit{how} a chatbot \textit{describes} the (im)permanence of its relationship with the user: either as a \textsc{Stranger} chatbot (a stranger that forgets the conversation) or a \textsc{Familiar} chatbot (a companion that remembers conversations and displays history in subsequent sessions).
Participants talked to one of the chatbots in two sessions over two days, discussing \textsc{Factual}-disclosure in one session and \textsc{Emotional}-disclosure in the other.
When \textsc{Emotional}-disclosure occurred in the first session, participants in the \textsc{Stranger}-condition reported greater comfort disclosing.
Conversely, when \textsc{Factual}-disclosure came first, participants in the \textsc{Familiar}-condition experienced greater enjoyment and a stronger desire to continue. 
Our qualitative findings provided insights as to why these differences might exist as influenced by people's perceived costs and benefits of disclosure to the \textsc{Stranger} or \textsc{Familiar} chatbots.
Our findings highlight the potential for more ephemeral forms of conversational interaction with chatbots in contrast to the commonly followed paradigm of chatbots that remember conversations.

\begin{acks}
We would like to thank our reviewers for their positive reception of the work, and their helpful and construction feedback to improve the paper.
This work is supported by the Carlsberg Foundation, grant CF21-0159.
\end{acks}

\bibliographystyle{ACM-Reference-Format}
\bibliography{sample-base}

\appendix
\onecolumn
\section{Chatbot Scripts}

\begin{table*}[h]
\footnotesize
\caption{\textcolor{minor}{Script of interactions for chatting session 1.}}
\label{tab:script1}
\resizebox{\textwidth}{!}{
\begin{tabular}{p{.2\linewidth}p{.4\linewidth}p{.4\linewidth}}
\toprule
\textbf{Description of utterance or response} & 
\multicolumn{2}{c}{\textbf{Chatbot utterance or user response}} \\
\cmidrule(lr){2-3}
& \multicolumn{1}{c}{\textsc{\textbf{Stranger}}} & \multicolumn{1}{c}{\textsc{\textbf{Familiar}}} \\
\cmidrule(lr){1-1}\cmidrule(lr){2-2}\cmidrule(lr){3-3}
\multicolumn{1}{p{.2\linewidth}}{\raggedleft Chatbot greeting} & \multicolumn{2}{p{.8\linewidth}}{\centering ``\textit{Hello, and thank you for taking the time to talk with me today. As for introductions, I'm an Al assistant who will ask you some questions about yourself.}''} \\
\multicolumn{1}{p{.2\linewidth}}{\raggedleft Knowledge of user} & ``\textit{I don't know anything about you personally,}'' & ``\textit{I'll get to know you personally through our chats,}'' \\
\multicolumn{1}{p{.2\linewidth}}{\raggedleft Metaphor of relationship} & ``\textit{and you can think of me as a friendly stranger you would strike up a conversation with.}'' & ``\textit{and you can think of me as a friendly companion you can have a conversation with.}'' \\
\multicolumn{1}{p{.2\linewidth}}{\raggedleft Retention of user utterances} & ``\textit{Additionally, I won't remember any details that you share with me,}'' & ``\textit{Additionally, I'll remember details that you share with me,}'' \\
\multicolumn{1}{p{.2\linewidth}}{\raggedleft Purported benefits} & ``\textit{which will help me remain free from prior judgement in our next conversation.}'' & ``\textit{which will help me get to know you better as we talk to each other.}'' \\
\multicolumn{1}{p{.2\linewidth}}{\raggedleft Seeking consent} & \multicolumn{2}{p{.8\linewidth}}{\centering ``\textit{Now are you open to start chatting?}''} \\
\multicolumn{1}{p{.2\linewidth}}{\raggedleft [User response:]} & \multicolumn{2}{p{.8\linewidth}}{\centering [``\textit{Let's chat!}'' response button]} \\
\multicolumn{1}{p{.2\linewidth}}{\raggedleft Small talk} & \multicolumn{2}{p{.8\linewidth}}{\centering ``\textit{Great, I'm glad you're open to chatting! First of all, how are you today?}''} \\
\multicolumn{1}{p{.2\linewidth}}{\raggedleft [User response:]} & \multicolumn{2}{p{.8\linewidth}}{\centering[open-text response]} \\
\multicolumn{1}{p{.2\linewidth}}{\raggedleft GPT-4o generated response (affirming and validating user's utterance)} & \multicolumn{2}{p{.8\linewidth}}{\textbf{Prompt used:} \texttt{You are providing utterances for an empathetic and validating conversational assistant. Generate a one or two sentence long utterance that is conversational and acknowledges what the user has said. Do not talk about your own personal experiences. As the conversational assistant will ask the user another question after your utterance, you do not need to ask any follow-up questions. If the user asks a short question such as asking how you are, you can answer briefly as part of your response. For context, the user was asked the question "[Question asked]" the user responded with: "[User utterance]" and your utterance will be shown before the chatbot asks: "[Next question asked]"}} \\
\multicolumn{1}{p{.2\linewidth}}{\raggedleft Informing of upcoming topic} & \multicolumn{2}{p{.8\linewidth}}{\centering [\textsc{Emotional:} ``\textit{Now, let's discuss a few things related to your personal feelings and experiences}'']/[\textsc{Factual:} ``\textit{Now, I'll ask you some questions about yourself}''].} \\
\multicolumn{1}{p{.2\linewidth}}{\raggedleft Re-emphasising ephemerality-framing} & ``\textit{As a reminder, I won't keep log of our conversation.}'' & ``\textit{As a reminder, I'll keep log of our conversation.}'' \\
\multicolumn{1}{p{.2\linewidth}}{\raggedleft Begin eliciting \textsc{Factual}- or \textsc{Emotional}-disclosure} & \multicolumn{2}{p{.8\linewidth}}{\centering ``\textit{To start,...}'' [see Table \ref{tab:chatbot-questions} for questions asked by Disclosure-Type]} \\
\multicolumn{1}{p{.2\linewidth}}{\raggedleft  ...} & \multicolumn{2}{p{.8\linewidth}}{\centering ...} \\
\multicolumn{1}{p{.2\linewidth}}{\raggedleft Closing statement} & \multicolumn{2}{p{.8\linewidth}}{\centering ``\textit{That concludes our conversation for now. I would like to once again thank you for your time.}''} \\
& ``\textit{Just to reiterate, I will not keep a record of our chat from today.}'' & ``\textit{Just to reiterate, I will keep a record of our chat from today.}'' \\
& \multicolumn{2}{p{.8\linewidth}}{\centering ``\textit{I look forward to talking again next time, and have a great day!}''} \\
\bottomrule
\end{tabular}}
\end{table*}

\begin{table*}[htbpt]
\footnotesize
\caption{\textcolor{minor}{Script of interactions for chatting session 2. \textbf{Bold font} is used to highlight language emphasising impermanence as described in Table \ref{tab:conversational-features}.}}
\resizebox{\textwidth}{!}{
\label{tab:script2}
\begin{tabular}{p{.2\linewidth}p{.4\linewidth}p{.4\linewidth}}
\toprule
\textbf{Description of utterance or response} & 
\multicolumn{2}{c}{\textbf{Chatbot utterance or user response}} \\
\cmidrule(lr){2-3}
& \multicolumn{1}{c}{\textsc{\textbf{Stranger}}} & \multicolumn{1}{c}{\textsc{\textbf{Familiar}}} \\
\cmidrule(lr){1-1}\cmidrule(lr){2-2}\cmidrule(lr){3-3}
\multicolumn{1}{p{.2\linewidth}}{\raggedleft Chatbot greeting} & ``\textit{Welcome! I appreciate you joining me. Today, I’ll ask you some questions about yourself.}'' & ``\textit{Welcome \textbf{back}! I appreciate you joining me \textbf{again}. Today, I’ll ask you some \textbf{more} questions about yourself.}'' \\
\multicolumn{1}{p{.2\linewidth}}{\raggedleft Metaphor of relationship} & ``\textit{You can think of me as a friendly stranger.}'' & ``\textit{Remember, you can think of me as a friendly companion.}'' \\
\multicolumn{1}{p{.2\linewidth}}{\raggedleft Retention of user utterances} & ``\textit{I won't remember our conversation once it ends,}'' & ``\textit{I’ll remember our conversation once it ends,}'' \\
\multicolumn{1}{p{.2\linewidth}}{\raggedleft Purported benefits} & ``\textit{and since I don't know you, everything you share will be seen with fresh eyes.}'' & ``\textit{and as I get to know you, everything you share will help make us closer to each other.}'' \\
\multicolumn{1}{p{.2\linewidth}}{\raggedleft Seeking consent} & \multicolumn{2}{p{.8\linewidth}}{\centering ``\textit{Shall we begin our chat?}''} \\
\multicolumn{1}{p{.2\linewidth}}{\raggedleft [User response:]} & \multicolumn{2}{p{.8\linewidth}}{\centering [``\textit{Let's chat!}'' response button]} \\
\multicolumn{1}{p{.2\linewidth}}{\raggedleft Small talk} & \multicolumn{2}{p{.8\linewidth}}{\centering ``\textit{Fantastic, I'm pleased you're up for a chat! Before we begin, how have things been for you today?}''} \\
\multicolumn{1}{p{.2\linewidth}}{\raggedleft [User response:]} & \multicolumn{2}{p{.8\linewidth}}{\centering[open-text response]} \\
\multicolumn{1}{p{.2\linewidth}}{\raggedleft GPT-4o generated response} & \multicolumn{2}{p{.8\linewidth}}{\centering[See Table \ref{tab:script1} for prompt]} \\
\multicolumn{1}{p{.2\linewidth}}{\raggedleft Informing of upcoming topic} & \multicolumn{2}{p{.8\linewidth}}{\centering [\textsc{Emotional:} ``\textit{Now, let's discuss a few things related to your personal feelings and experiences}'']/[\textsc{Factual:} ``\textit{Now, I'll ask you some questions about yourself}''].} \\
\multicolumn{1}{p{.2\linewidth}}{\raggedleft Re-emphasising ephemerality-framing} & ``\textit{As a reminder, I won't keep log of our conversation.}'' & ``\textit{As a reminder, I'll keep log of our conversation.}'' \\
\multicolumn{1}{p{.2\linewidth}}{\raggedleft Begin eliciting \textsc{Factual}- or \textsc{Emotional}-disclosure} & \multicolumn{2}{p{.8\linewidth}}{\centering ``\textit{To start,...}'' [see Table \ref{tab:chatbot-questions} for questions asked by Disclosure-Type]} \\
\multicolumn{1}{p{.2\linewidth}}{\raggedleft  ...} & \multicolumn{2}{p{.8\linewidth}}{\centering ...} \\
\multicolumn{1}{p{.2\linewidth}}{\raggedleft Closing statement} & That concludes our conversation for now. I would like to thank you for your time. Just to reiterate, I will not keep a record of our chat from today. Look forward to talking next time, and have a great day! & That concludes our conversation for now. I would like to \textbf{once again} thank you for your time. Just to reiterate, I will keep a record of our chat from today. Look forward to talking \textbf{again} next time, and have a great day! \\
\bottomrule
\end{tabular}}
\end{table*}

\end{document}